\documentclass[twocolumn]{aastex701}
\usepackage{amsmath}
\usepackage{lipsum}
\usepackage{booktabs}
\usepackage{makecell}
\usepackage{afterpage}
\usepackage{listings}
\usepackage{xcolor}
\usepackage{fancyvrb}
\usepackage{fvextra}

\definecolor{codegreen}{rgb}{0,0.6,0}
\definecolor{codegray}{rgb}{0.25,0.25,0.25}
\definecolor{codepurple}{rgb}{0.38,0,0.82}
\definecolor{backcolour}{rgb}{0.975,0.975,0.975}

\lstdefinelanguage{http}{%
    sensitive=t,%
    alsoletter=-,%
    morekeywords={cache-control,pragma,transfer-encoding,content-type,expires,server,x-aspnet-version,x-powered-by,access-control-allow-origin,access-control-allow-headers,access-control-allow-methods,date,strict-transport-security,content-security-policy,set-cookie},%
}
\lstdefinestyle{mystyle}{
    morekeywords={access,and,break,class,continue,def,del,elif,else,except,exec,
    finally,for,from,global,if,import,in,is,lambda,not,or,pass,print,raise,
    return,self,try,while},
    morekeywords=[2]{abs,all,any,basestring,bin,bool,bytearray,callable,chr,
    classmethod,cmp,compile,complex,delattr,dict,dir,divmod,enumerate,eval,
    execfile,file,filter,float,format,frozenset,getattr,globals,hasattr,hash,
    help,hex,id,input,int,isinstance,issubclass,iter,len,list,locals,long,map,
    max,memoryview,min,next,object,oct,open,ord,pow,property,range,raw_input,
    reduce,reload,repr,reversed,round,set,setattr,slice,sorted,staticmethod,str,
    sum,super,tuple,type,unichr,unicode,vars,xrange,zip,apply,buffer,coerce,
    intern},
    backgroundcolor=\color{white},
    commentstyle=\color{codegreen},
    keywordstyle=\ttfamily\color{codepurple},
    numberstyle=\footnotesize\ttfamily\color{codegray},
    stringstyle=\color{codepurple},
    basicstyle=\small\ttfamily,
    breakatwhitespace=false,
    breaklines=true,
    captionpos=b,
    keepspaces=true,
    numbers=left,
    numbersep=10pt,
    showspaces=false,
    showstringspaces=false,
    showtabs=false,
    tabsize=2,
    frame=Tb,
    frameround={t}{t}{t}{t},
    framexleftmargin=7mm,
    xleftmargin=22pt,
    xrightmargin=3pt
    }
\graphicspath{{./}{./figures/}}

\newcommand{\re}{$r_e$}
\newcommand{\magr}{$r_\mathrm{mag}$}
\newcommand{\virial}{$R_{200}$}
\newcommand{\virialf}{$5R_{200}$}

\begin{document}

\title{AstroInspect: a web-based system to organize, assess, and visually inspect  astronomical objects}

\correspondingauthor{Natanael M. Cardoso}

\author[orcid=0000-0002-2238-9665]{Natanael M. Cardoso}
\affiliation{Escola Politécnica, Universidade de São Paulo, São Paulo, 05508-010, SP, Brasil}
\email[show]{natanael.mc@usp.br}

\author[orcid=0000-0002-5267-9065]{Claudia Mendes de Oliveira} 
\affiliation{Instituto de Astronomia, Geofísica e Ciências Atmosféricas, Universidade de São Paulo, São Paulo, 05508-090, SP, Brasil}
\email{claudia.oliveira@iag.usp.br}

\author[orcid=0000-0003-4630-1311]{Angela C. Krabbe}
\affiliation{Instituto de Astronomia, Geofísica e Ciências Atmosféricas, Universidade de São Paulo, São Paulo, 05508-090, SP, Brasil}
\email{angela.krabbe@gmail.com}

\author[orcid=0009-0007-2396-0003]{Analía V. Smith Castelli}
\affiliation{Instituto de Astrofísica de La Plata, CONICET-UNLP, La Plata, B1900FWA, LP, Argentina.}
\affiliation{Facultad de Ciencias Astronómicas y Geofísicas, Universidad Nacional de La Plata, La Plata, B1900FWA, LP, Argentina}
\email{a.smith.castelli@gmail.com}

\author[orcid=0009-0003-6609-1582]{Gustavo B. Oliveira Schwarz}
\affiliation{Escola Politécnica, Universidade de São Paulo, São Paulo, 05508-010, SP, Brasil}
\email{gustavo.b.schwarz@gmail.com}

\author[orcid=0000-0001-6480-1155]{Lilianne Nakazono}
\affiliation{Observatório Nacional, Rio de Janeiro, 20921-400, RJ, Brasil.}
\affiliation{Intituto de Física, Universidade de São Paulo, São Paulo, 05508-090, SP, Brasil}
\email{lilianne.nakazono@usp.br}

\author[orcid=0000-0003-3921-2177]{Ricardo Demarco}
\affiliation{Institute of Astrophysics, Facultad de Ciencias Exactas, Universidad Andr\'es Bello, Sede Concepci\'on, Talcahuano, Chile}
\email{demarco.rj@gmail.com}

\author[orcid=0009-0008-9042-4478]{Maiara S. Carvalho}
\affiliation{Instituto de Astronomia, Geofísica e Ciências Atmosféricas, Universidade de São Paulo, São Paulo, 05508-090, SP, Brasil}
\email{mscarvalho.astro@gmail.com}

\author[orcid=0000-0002-4064-7234]{William Schoenell}
\affiliation{The Observatories of the Carnegie Institution for Science, 813 Santa Barbara St, Pasadena, CA 91101, USA}
\email{wschoenell@carnegiescience.edu}

\author[orcid=0000-0002-0138-1365]{Tiago Ribeiro}
\affiliation{Rubin Observatory Project Office, Tucson, AZ 85719, USA}
\email{tribeiro@lsst.org}

\author[orcid=0009-0007-8005-4541]{Antonio Kanaan}
\affiliation{Departamento de Física, Universidade Federal de Santa Catarina, Florianópolis, 88040-900, SC, Brasil}
\email{ankanaan@gmail.com}

\author[orcid=0000-0003-2283-1123]{Antonio M. Saraiva}
\affiliation{Escola Politécnica, Universidade de São Paulo, São Paulo, 05508-010, SP, Brasil}
\affiliation{Center for Artificial Intelligence, Universidade de São Paulo, São Paulo, 05508-020, SP, Brasil}
\affiliation{Instituto de Estudos Avançados, Universidade de São Paulo, São Paulo, 05508-050, SP, Brasil}
\email{saraiva@usp.br}

\begin{abstract}
The rapid growth of imaging and spectroscopic surveys has intensified the need for efficient tools that support visual inspection, a practice that remains essential for tasks such as classification, catalog refinement, and validation of automated methods. Existing solutions, however, often require the use of multiple platforms and complex workflows to integrate heterogeneous data. To address this challenge, we present the first release of the AstroInspect (\url{https://astroinspect.github.io}), a web-based system which ensures seamless access to several astronomical resources. The system provides an intuitive graphical user interface (GUI) through which users can upload catalogs of objects defined by celestial coordinates. AstroInspect automatically enriches these catalogs with complementary information, including imaging, spectroscopic, and photometric data retrieved in real time from surveys such as the Sloan Digital Sky Survey (SDSS), the Legacy Surveys (LS), and the Southern Photometric Local Universe Survey (S-PLUS). As an example of its scientific utility, we used AstroInspect to identify H$\alpha$ emission-line galaxies within a 7 deg radius in the direction of the Hydra I cluster (also known as Abell 1060) by visual inspection. Using a candidate set of 981 galaxies selected from S-PLUS photometric data, we produced a catalog of 80 galaxies with confirmed H$\alpha$ emission. These results highlight the potential of AstroInspect to support efficient visual inspection workflows.
\end{abstract}


\section{Introduction}
\label{sec:introduction}
\begin{figure*}[t!]
    \centering
    \includegraphics[width=\textwidth]{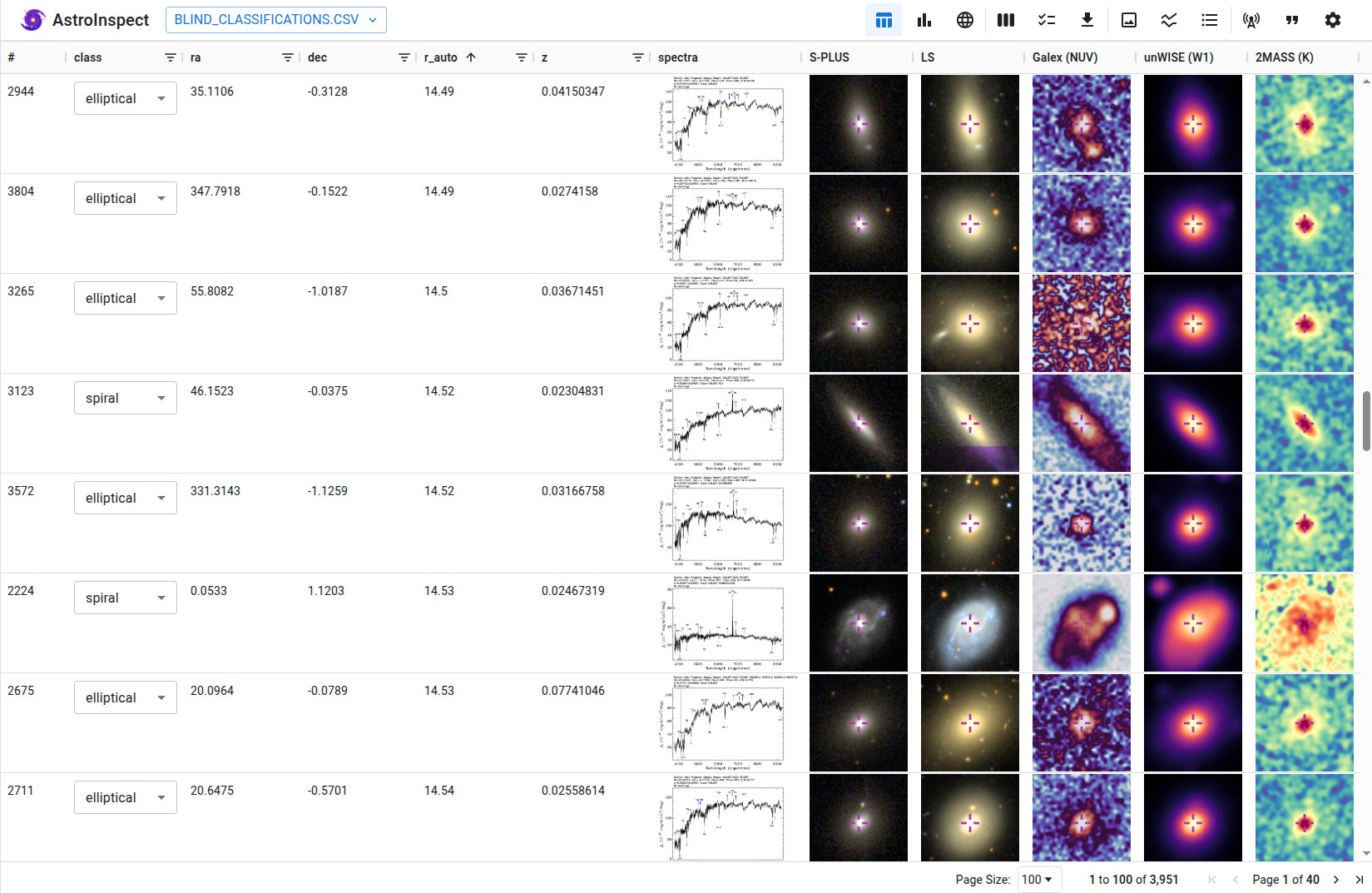}
    \caption{The AstroInspect interface streamlines visual inspection and scientific analysis by integrating spectroscopic, photometric, and imaging data into a single view, allowing ``on-the-fly''  comparative analyses. The user-uploaded catalog is organized into an interactive tabular interface that merges the input data (e.g. ``ra'', ``dec'', and ``r\_auto'' columns) with dynamically-obtained data from remote services, such as redshift (e.g. ``z'' column from the SDSS spectroscopic catalog), the spectral energy distribution (``spectra'' column), and color images from different surveys (e.g. ``S-PLUS'', ``LS'', ``Galex (NUV)'', ``unWISE (W1)'', and ``2MASS (K)'' columns, respectively). To support more effective visual assessment, AstroInspect also enables customization of image rendering parameters, including color stretch functions, colormap selection, and field-of-view adjustments.}
    \label{fig:screen}
\end{figure*}

Visual inspection of astronomical objects remains a fundamental step in observational astronomy, serving purposes such as morphological analysis \citep{nair2010,darg2010b,efigi,kartaltepe2015,paudel2018,astromorphlib,alexander2023,galdeano2025}, sample cleaning \citep{lambas2012,analia}, and validation of automated methods \citep{alexander2023,tan2025,clecio-1,natanael}. This process leverages human sensitivity to visual patterns in images alongside an understanding of the underlying physical context, which remains challenging to fully encode in algorithms.

Several tools have been developed for the access, analysis, and manipulation of astronomical data, addressing tasks such as the processing of tabular data, visualization of celestial images and atlases, spectral analysis, and the exploration of catalogs in remote databases.

For tabular data, the Starlink Tables Infrastructure Library Tool Set (STILTS; \citealp{stilts}) is a Java library for processing, cross-match, and statistical analysis of large catalogs, also accessible via a Command Line Interface (CLI). Its companion tool, the Tool for Operations on Catalogues and Tables (TOPCAT; \citealp{topcat, topcat-stil}), provides a widely used Graphical User Interface (GUI) for interactive visualization and manipulation of tabular data.

For spectral analysis, Specview \citep{specview, specview-2} and the Virtual Observatory (VO)  Spectral Analysis Tool (SPLAT-VO; \citealp{splat, splatvo, splatvo-timeseries}) offer GUI environments for the inspection and modeling of spectroscopic data.

In imaging, SAOImage DS9 \citep{saods9} is a tool for visualizing astronomical images, supporting multidimensional data handling, region definition, and image overlays. Aladin \citep{aladin, aladin-2, aladin-desktop} integrates sky surveys and catalogs, enabling visualization and exploration of sky atlases. Their web counterparts, SAOImage JS9 \citep{js9} and Aladin Lite \citep{aladin-lite}, provide access to the  core functionalities in web browsers.

In this work, we present the first release of the AstroInspect (\url{https://astroinspect.github.io}), a web-based system designed to address the gap related to softwares specialized in visual inspection and classification in the ecosystem of astronomical tools. The system integrates functionalities that enhances the efficiency of the workflow for morphological studies and exploratory analyses of astronomical catalogs by integrating data retrieval and visualization capabilities into a unified GUI (shown in Fig.~\ref{fig:screen}). AstroInspect enables users to upload object defined by celestial coordinates, right ascension (RA) and declination (Dec). The application then automatically retrieves and displays associated imaging, spectroscopic, and photometric data. These data products may originate from both pre-configured major surveys and user-provided datasets, supporting various analytical workflows.

To illustrate its scientific application, we employed AstroInspect to identify galaxies with H$\alpha$ emission line in the direction of the \object{Hydra I} galaxy cluster (also known as \object{Abell 1060}, \citealp{abell-clusters}). Candidate galaxies were first selected from S-PLUS photometric data and subsequently examined through visual inspection in AstroInspect. This analysis yielded a catalog of 80 galaxies with confirmed H$\alpha$ emission-line within an area with a projected radius of 7 deg, centered on the coordinates of Hydra I, providing a foundation for further studies and demonstrating the utility of the described tool.
 
In Section \ref{sec:data}, we describe the data sources used by AstroInspect. In Section \ref{sec:methods}, we detail the system design and implementation. In Section \ref{sec:features}, we summarize the system features. In Section \ref{sec:sci}, we present a scientific case study using AstroInspect to investigate H$\alpha$ emission candidates in the direction of  Hydra I cluster. In Section \ref{sec:discussion}, we discuss our findings. Finally, in Section \ref{sec:conclusion}, we present the conclusion and future work perspectives.

\section{Data Sources}
\label{sec:data}
The core feature of the AstroInspect is its ability to aggregate heterogeneous astronomical data within a unified graphical interface. This section summarizes the principal data sources integrated into the tool, outlining the distinct imaging, photometric, and spectroscopic products they provide.

\subsection{The Large Sky Area Multi-Object Fiber Spectroscopic Telescope}
\label{sec:lamost}
The Large Sky Area Multi-Object Fiber Spectroscopic Telescope (LAMOST; \citealp{lamost,lamost-overview}), also known as the Guoshoujing Telescope, is a facility located at the Xinglong Station of the National Astronomical Observatories, China. Designed as a quasi-meridian reflecting Schmidt telescope with an effective aperture of 4 meters, LAMOST is optimized for wide-field spectroscopic observations and is equipped with 4,000 fibers on its focal plane. With over a decade of operation, LAMOST obtained spectra from more than 25 million spectra, covering a footprint of $\sim$20,000 deg$^2$ of the northern sky. These spectroscopic data is obtained through the LAMOST public   API\footnote{\url{https://lamost.org/openapi/docs}}.

\subsection{The Dark Energy Spectroscopic Instrument}
\label{sec:desi}
The Dark Energy Spectroscopic Instrument (DESI; \citealp{desi}) is a wide-field optical spectrograph with a focal plane equipped with 5,000 robotically controlled fiber positioners installed on the 4-meter Mayall Telescope at Kitt Peak National Observatory. The ongoing five-year survey will obtain spectra and redshifts for tens of millions of galaxies, stars, and quasars across approximately 14,000 deg$^2$ of the northern sky.

The DESI spectra is obtained through SPectra Analysis and Retrievable Catalog Lab (SPARCL; \citealp{sparcl}) API\footnote{\url{https://astrosparcl.datalab.noirlab.edu}} available at the NOIRLab Astro Data Lab science platform\footnote{\url{https://datalab.noirlab.edu}} \citep{datalab2019,datalab}.

\subsection{The Sloan Digital Sky Survey}
\label{sec:sdss}
The Sloan Digital Sky Survey\footnote{\url{https://sdss.org}} (SDSS; \citealp{sdss}) is an imaging and spectroscopic survey initiated the year 2000 employing a 2.5-meter telescope \citep{sdss-telescope} to map the northern sky across five bands, $u$, $g$, $r$, $i$, and $z$ \citep{sdss-filters}. The SDSS spectrograph \citep{sdss-spec} obtained measurements of the Spectral Energy Distribution (SED) for more than 5 million objects of the northern sky. 

Both photometric and spectroscopic data are obtained through the SkyServer\footnote{\url{https://skyserver.sdss.org}} \citep{skyserver, sciserver} Application Programming Interface (API).

\subsection{The Southern Photometric Local Universe Survey}
\label{sec:splus}
The Southern Photometric Local Universe Survey\footnote{\url{https://www.splus.iag.usp.br}} (S-PLUS; \citealp{splus}) is an imaging survey that covers a large portion of the southern sky using a wide-field telescope, the T80-South, located in Chile. It employs a 12-band photometric system that includes five broadband filters similar to those of the SDSS and seven narrow-band filters, selected to cover key spectral features. The narrow-band filters of S-PLUS are centered on the Balmer jump/[O{\sc ii}] (also known as F378), Ca H + K (F395), H$\delta$ (F410), G band (F430), Mg b triplet (F515), H$\alpha$ (F660), and Ca triplet (F861) (see Table 2 in \citealp{splus}). The filters F395, F410, and F430 cover the H$\epsilon$, H$\delta$, and H$\gamma$ lines, respectively. This wide spectral coverage, ranging from ultraviolet to near-infrared, enables the construction of photometric SEDs (also called photo-spectra), which are particularly valuable when spectroscopic SEDs (also called spectra) are unavailable. 

The imaging data, reduced by the Multiband Astronomical Reduction (MAR) package \citep{mar}, and photometric catalog from S-PLUS are retrieved from the S-PLUS Cloud API\footnote{\url{https://splus.cloud}}. 

\subsection{DESI Legacy Imaging Surveys}
\label{sec:legacy}
The DESI Legacy Imaging Surveys\footnote{\url{https://legacysurvey.org}} (LS; \citealp{legacy}) are a collaborative effort comprising three public projects: the Dark Energy Camera (DECam) Legacy Survey (DECaLS; \citealp{decals}), the Beijing-Arizona Sky Survey (BASS; \citealp{bass}), and the Mayall $z$-band Legacy Survey (MzLS; \citealp{mzls}). Together, these surveys cover both the northern and southern hemispheres, capturing images of the extragalactic sky  in four optical bands ($g$, $r$, $i$, and $z$) using telescopes at the Kitt Peak National Observatory and Cerro Tololo Inter-American Observatory. 

The imaging data is retrieved from the Legacy Survey Sky Browser API\footnote{\url{https://legacysurvey.org/viewer}} and the photometric catalog is obtained through the NOIRLab Astro Data Lab science platform.

\subsection{Hierarchical Progressive Surveys}
\label{sec:hips}
In addition with the previously mentioned data sources, AstroInspect also integrates imaging information from several Hierarchical Progressive Surveys (HiPS; \citealp{hips}) covering a wide spectral range. The available surveys are the Galaxy Evolution Explorer (Galex; \citealp{galex}), the SDSS, the Skymapper \citep{skymapper-dr4}, the Panoramic Survey Telescope and Rapid Response System (Pan-STARRS; \citealp{panstarrs}), the Dark Energy Survey (DES; \citealp{des}), the Hyper Suprime-Cam Subaru Strategic Program (HSC; \citealp{hsc-dr1}), the Deep Near-Infrared Southern Sky Survey (DENIS; \citealp{denis}), the Two Micron All Sky Survey (2MASS; \citealp{twomass}), the United Kingdom Infrared Telescope (UKIRT; \citealp{ukirt-dr1}) Infrared Deep Sky Survey (UKIDSS; \citealp{ukidss}), the UKIRT Hemisphere Survey (UHS; \citealp{uhs-dr1}), the Unblurred Wide-field Infrared Survey Explorer (unWISE; \citealp{wise,unwise}), and the Karl G. Jansky Very Large Array Sky Survey (VLASS; \citealp{VLASS}). For redundancy purposes, AstroInspect also includes HiPS for the LS.

The imaging data is retrieved from Hips2Fits\footnote{\url{https://alasky.cds.unistra.fr/hips-image-services/hips2fits}} service provided by the Centre de Données Astronomiques de Strasbourg\footnote{\url{https://cds.unistra.fr}} (CDS).

\section{System design and implementation}
\label{sec:methods}

\subsection{System overview and design concept}
\label{sec:design-overview}
Visual classification studies that involve inspecting large astronomical catalogs frequently begin with the development of dedicated graphical tools to support the classification process (e.g., \citealp{nair2010,efigi,kartaltepe2015,boyce2023}). To support future research with a flexible and reusable solution, we have developed AstroInspect with the following key features:

\begin{enumerate}
    \item \textbf{System generality and extensibility:} the system shall support multiple scientific workflows and shall allow the integration of both externally published survey data and user-provided local datasets.
    \item \textbf{Automated data acquisition:} the system shall provide configurable mechanisms for the automatic retrieval of data from multiple external services.
    \item \textbf{Performance and concurrency:} the system shall employ concurrent data-access strategies to reduce latency and maintain responsiveness when processing large catalogs.
    \item \textbf{Multi-survey data integration:} the system shall combine data products from multiple astronomical surveys to maximize spatial and spectral coverage for each object.
    \item \textbf{Multi-modal data visualization:} the system shall present imaging and spectroscopic data concurrently to support accurate visual inspection and classification tasks.
    \item \textbf{Object classification and annotation:} the system shall provide functionality for users to assign classifications and annotations to astronomical objects directly within the input table and shall store these results for subsequent analysis or export.
    \item \textbf{Usability and accessibility:} the system shall provide an intuitive user interface that does not require programming expertise, virtual environment configuration, or software installation.
\end{enumerate}

Based on these functional requirements, we designed AstroInspect to provide an integrated environment for the inspection, visualization, and classification of astronomical catalogs through a streamlined and accessible workflow, as shown in Fig. \ref{fig:screen}. The user initiates the process by loading an input table, which can be stored locally or accessed remotely. For each row (representing a single astronomical object), the application automatically queries remote services based on the object's coordinates (RA, Dec). The retrieved material augments the input catalog with imaging, spectroscopic, or catalog data products, depending on the user's configuration. To ensure broad applicability, the tool also supports the display of arbitrary local or remote datasets supplied by the user. 

Furthermore, the system includes native annotation functionality, enabling users to record classifications or comments directly within the interface. The architecture that enables this workflow comprises several interconnected components, described in Section \ref{sec:comp}.

\subsection{System components}
\label{sec:comp}
This section provides a high-level description of the system's core components, which are responsible for orchestrating data retrieval, state management, and user interaction. The overall architecture diagram is shown in Fig. \ref{fig:arch}.

\subsubsection{Table State Manager}
\label{sec:comp-state}
The Table State Manager (TSM) maintains the application's global state throughout execution, providing a centralized interface for accessing and modifying system data. It stores the user-loaded table, retrieved survey products, and all user interactions, including configuration choices and annotations. All system components interact with the TSM to read or update the current state, ensuring consistent data flow and coordinated behavior across the application.

\subsubsection{Table I/O Handler}
\label{sec:comp-table-handler}
The Table Input/Output (I/O) Handler (TIOH) manages the import and export of tabular data and extracts essential metadata, such as the names of coordinate columns. During data loading, the TIOH parses the input table and dispatches the extracted content in internal representation to the TSM for storage. When exporting, it retrieves the current table state from the TSM  (including user-driven modifications such as filtering, reordering, or sorting) and generates a downloadable file for the user.

\subsubsection{Tasks}
\label{sec:comp-workers}
Tasks refer to all operations that retrieve remote resources or process acquired data for display in the GUI. Tasks are classified as either compute-bound or I/O-bound.

\begin{figure*}[p]
    \centering
    \includegraphics[width=\linewidth]{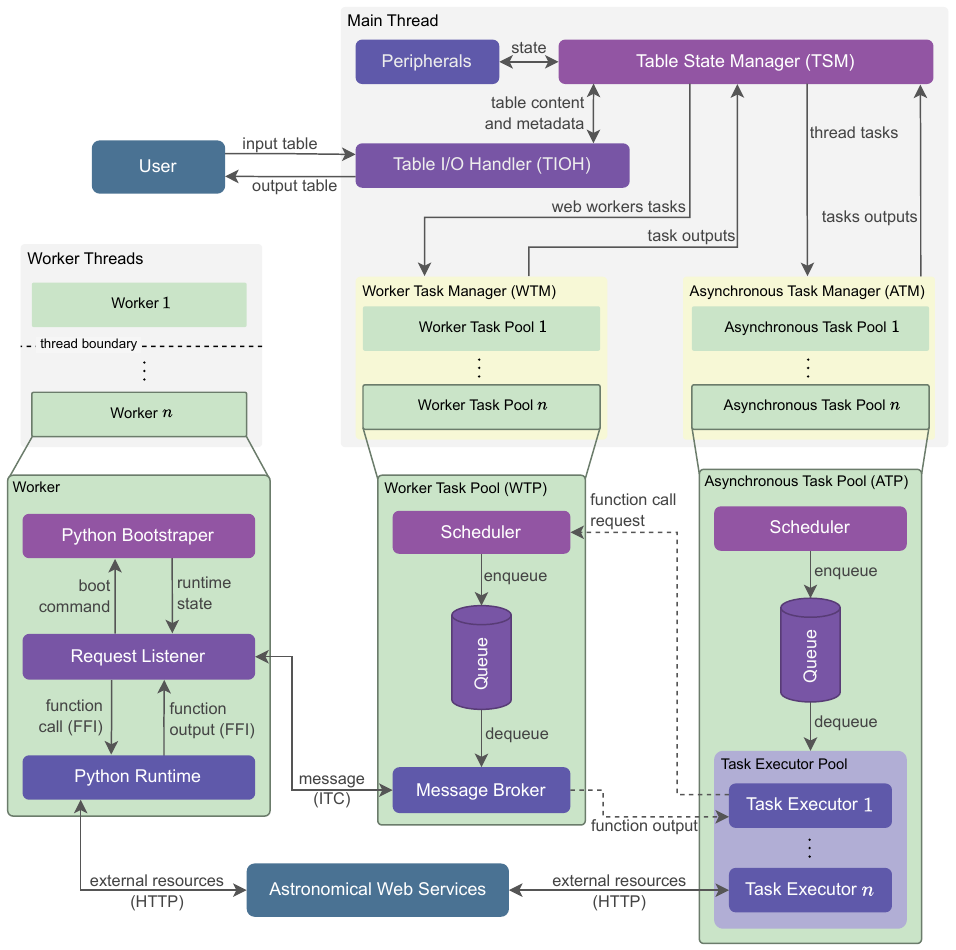}
    \caption{A schematic of the AstroInspect system architecture, illustrating the interaction of its core components. The Table State Manager (TSM) maintains the global application state, and the Table I/O Handler (TIOH) manages data import and export. Concurrent task execution is managed by two specialized pools: the Worker Task Pool (WTP), which orchestrates computationally intensive tasks in isolated threads via workers running Python code, and the Asynchronous Task Pool (ATP), which handles I/O tasks via non-blocking Javascript functions running in main thread. The Request Listener of the Worker interacts with the Python Runtime calling functions and receiving their results back via Foreign Function Interface (FFI), which enables data exchange between the JavaScript and Python. The Message Broker facilitates Inter-Thread Communication (ITC) between workers and the main thread. The executors (workers or asynchronous functions) can retrieve resources from Astronomical Web Services through Hypertext Transfer Protocol (HTTP). After a task is completed, its results (output data or error message) are dispatched to the TSM, being able to be displayied in the GUI. Peripheral components interact with the TSM based on user actions. Solid arrows represent the data flow between components, while dashed arrows shows an optional relation between ATP and WTP forming hybrid tasks, in which the ATP initiates the task by performing I/O operations, forwards processing workloads to the WTP, and receives the result back, following a client-server relationship.}
    \label{fig:arch}
\end{figure*}

Compute-bound tasks require local processing prior to visualization (e.g., generating photometric SEDs from S-PLUS magnitudes). These tasks are handled by the Worker component (Fig. \ref{fig:arch}), which executes Python code in isolated threads via the browser's Web Worker API\footnote{\url{https://html.spec.whatwg.org/multipage/workers.html}}. This prevents computationally intensive operations from blocking the responsiveness of the GUI. Each Worker contains three internal modules: the Python Bootstrapper, which initializes the interpreter and the required packages; the Python Runtime, which executes python code related to  HTTP-based data retrieval and processing; and the Request Listener, which manages communication between the Python interpreter via Foreign Function Interface (FFI) and the main thread via Inter-Thread Communication (ITC).

Conversely, I/O-bound tasks primarily involve data transfer with minimal processing (e.g., displaying pre-cut images obtained from remote services). These tasks execute on the main thread using non-blocking JavaScript functions managed by the browser event loop. 

This design separates computational and I/O workloads to optimize performance. Workers enable parallel execution of compute-intensive tasks but incur overhead from thread initialization, Python runtime startup, transformation of the memory representation of objects between Python and Javascript runtimes (called marshaling), and transport across ITC boundaries. Asynchronous functions avoid these costs by sharing the memory space with the main application thread, but they are unsuitable for long computation tasks.

\subsubsection{Task Pools}
\label{sec:comp-task-pools}
The task pools limits the number of concurrent operations executed for each data provider. AstroInspect implements two task pools: the Worker Task Pool (WTP) for compute-bound tasks and the Asynchronous Task Pool (ATP) for I/O-bound tasks. This separation provides fine-grained control over concurrency, reduces load on external services, and prevents processing bottlenecks. It also enables hybrid execution workflows, in which tasks begin in the ATP for remote data acquisition and continue in the WTP for local processing, as indicated by the dashed arrows in Fig. \ref{fig:arch}.

Both task pools follow a producer–consumer model \citep{producer-consumer68,producer-consumer72}. The Scheduler component acts as the producer, generating and queuing task messages that specify the function to be executed, its arguments, and the preassigned executor within the pool. Task assignment follows a round-robin scheduling strategy.

In the WTP, workers are managed through a preallocated pool to avoid the overhead of dynamic thread creation. The Message Broker acts as the consumer, dequeuing tasks and dispatching them to the worker selected by the Scheduler via Inter-Thread Communication (ITC). Upon creation, each worker's Request Listener automatically subscribes to the Message Broker, allowing it to receive function-call requests and return execution results through the communication same channel.

In contrast, the ATP uses a Task Executor Pool as the consumer. It executes asynchronous, non-blocking functions on the main thread controled by browser's event loop.

\subsubsection{Task Managers}
\label{sec:comp-task-managers}
The system employs high-level task manager components to orchestrate concurrency. AstroInspect implements two managers: the Worker Task Manager (WTM) for thread-based pools (WTPs) and the Asynchronous Task Manager (ATM) for non-blocking function pools (ATPs). Each manager is responsible to create, control, and terminate dedicated task pool instances responsible for all data retrieval and processing tasks from a specific web service, such as SkyServer, S-PLUS Cloud, or Astro Data Lab.

\subsubsection{Peripherals}
\label{sec:comp-pheripherials}
The peripheral components include all system modules not directly involved in concurrent task orchestration and therefore not shown in Fig. \ref{fig:arch}. These elements, such as configuration forms and interface utilities, interact with the TSM to read the current application state and dispatch updates when user actions or settings require changes. Their role is to support user interaction while maintaining consistency across the interface.

\begin{figure}[b!]
    \centering
    \includegraphics[width=\linewidth]{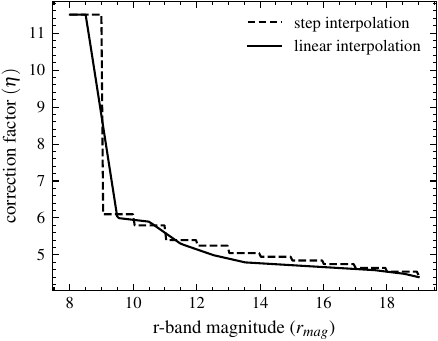}
    \caption{Empirical correction factor ($\eta$) as a function of the $r$-band magnitude (\magr) used to estimate the optimal field of view (FOV) for the stamps. Two interpolation schemes were employed to represent $\eta$: step interpolation (rank 0, dashed line) and linear interpolation (rank 1, solid line). The linear interpolation was implemented in AstroInspect, as it provides a better fit for brighter objects.}
    \label{fig:correction_factor}
\end{figure}

\begin{figure*}[t!]
    \centering
    \includegraphics[width=\textwidth]{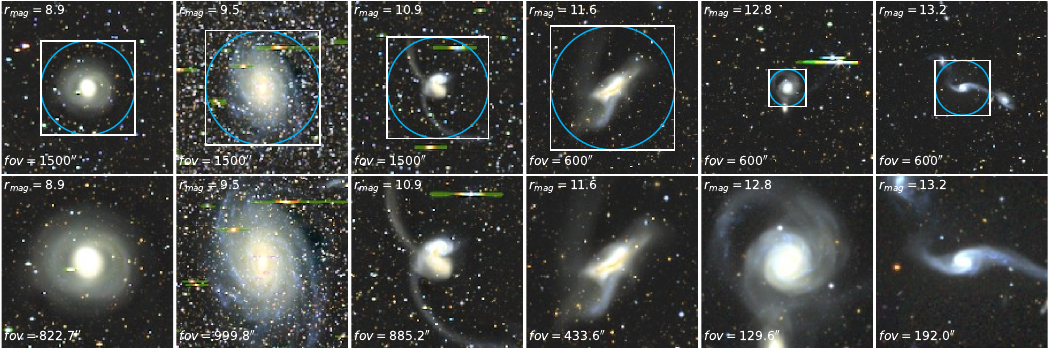}\\[1pt]
    \includegraphics[width=\textwidth]{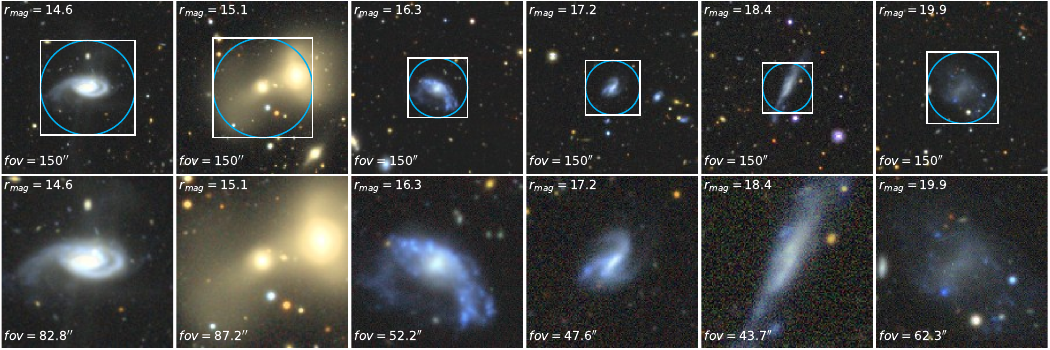}
    \caption{Adjustment of the LS image cutouts for twelve galaxies. The upper two rows show objects with $r$-band magnitude range from 8 to 14 mag, while the lower two rows show objects with $r$-band magnitude range from 14 to 20 mag. In the first and third rows, the panels illustrate the initial fields of view (FOVs) manually defined prior to the visual inspection procedure described in Section \ref{sec:stamps}. The blue circle marks the effective radius (\re), scaled according to the Eq. \ref{eq:fov}. The white square indicates the bounding box associated with this circle. The second and fourth rows present the final automatically adjusted cutouts, whose FOVs correspond to the bounding boxes shown above.}
    \label{fig:stamps-fitting-1}
\end{figure*}

\subsection{Implementation details}
We adopt a reactive programming paradigm to coordinate data flow among system components under high data-throughput conditions. In this model, the interface updates automatically in response to changes in the application state (managed by the TSM, see Fig. \ref{fig:arch}), such as the completion of remote data requests or user interactions. This approach ensures that data products appear in the GUI as soon as they become available, minimizing the latency in the visual feedback and improving the user experience.

To support this architecture, the client interface is developed using Next.js\footnote{\url{https://nextjs.org}}, a framework built on React.js\footnote{\url{https://react.dev}} that supports reactive programming and static site generation. By serving the application as precompiled static assets, the system avoids server-side page rendering at runtime, reducing the backend complexity.

WebAssembly (WASM) is a portable binary format that enables high-performance execution of code from languages other than Javascript directly within the web browser. For this work, we leverage Pyodide, a distribution of the scientific Python ecosystem compiled to WASM. This provides client-side access to essential astronomical Python libraries without local installation, a capability increasingly adopted for scientific web applications (e.g., \citealp{torrens2024,harris2024,ji2024,heil2025,peeters2025,zakova2025}). This choice  directly fulfills the accessibility requirement central to our system design (Section \ref{sec:design-overview}), delivering an analysis environment that operates entirely within a web browser.

Python code is used to generate both photometric and spectroscopic SED plots from S-PLUS and DESI surveys, respectively, using the Matplotlib package \citep{matplotlib}, and to handle table I/O operations with Astropy package \citep{astropy2013, astropy2018, astropy2022}.

\subsection{Image cutout preparation}
\label{sec:stamps}
A key consideration when creating image cutouts is the angular field of view (FOV), which defines the extent of the sky covered. When dealing with heterogeneous datasets that include objects with a wide distribution of sizes and distances, a single fixed FOV for the entire dataset can complicate the inspection process. Smaller objects will be displayed in a very reduced form or, conversely, large objects will be presented in very enlarged sections, making it impossible to see their entire structure. Therefore, an automatically adjusted FOV ensures that each object can be visualized within its context and in relation to its morphological characteristics.

The effective radius (\re) was adopted as the basis for estimating the FOV of each object. This parameter, defined as the radius enclosing half of the total emitted light, is provided in the LS photometric catalog (Section \ref{sec:legacy}). To approximate the FOV, \re~must be scaled by a correction factor ($\eta$), which depends on the object's brightness, expressed by its $r$-band magnitude (\magr), as given in Eq. \ref{eq:fov}:

\begin{equation}\label{eq:fov}
    \mathrm{fov}(r_\mathrm{mag}, r_e) = 2 \eta(r_\mathrm{mag}) r_e
\end{equation}

The correction factor ($\eta$) was empirically derived through the visual inspection of 12 galaxy samples, each comprising 500 objects within intervals of 1 mag in the $r$ band, covering the brightness range from 8 to 20 mag. To perform this inspection, we employed AstroInspect's functionality for loading custom images (Section \ref{sec:interop}), which enabled the assessment of each sample and the determination of optimal correction factor values. Both step interpolation (spline of rank 0) and linear interpolation (spline of rank 1) were evaluated, and the resulting dependence of $\eta$ on \magr~is presented in Fig. \ref{fig:correction_factor}. We choose to use the linear interpolation in the AstroInspect implementation, as it fits better on brighter objects.

To illustrate the developed approach, Fig. \ref{fig:stamps-fitting-1} presents the automatic fitting results for a randomly selected galaxy from each of the 12 samples, corresponding to the magnitude intervals 8-20 mag.

AstroInspect retrieves adjusted image cutouts in three steps. First, the photometric data (\re~and \magr) of the object are retrieved by the VO service provided by Astro Data Lab using the Simple Cone Search (SCS; \citealp{scs}) protocol, parametrized with the RA and Dec coordinates of the object from the input table and a search radius of 1.5 arcsec. Then, the FOV is calculated using Eq. \eqref{eq:fov} with the retrieved \re~and \magr. Finally, a request is made to the image services to retrieve an image cutout within the calculated FOV.

\section{System features}
\label{sec:features}
In this section, we highlight the signature features available in AstroInspect.

\subsection{SDSS catalogs access}
\label{sec:external}
Cross-referencing observational data across multiple catalogs is often essential in astronomical research to fully characterize observed objects. AstroInspect enables real-time retrieval of columns from SDSS photometric and spectroscopic catalogs via the SkyServer API, allowing the display of measurements such as magnitudes, redshifts, and other parameters directly selected through the GUI.

\subsection{Image cutout visualization}
\label{sec:object-stamps}
The image cutout viewing feature in AstroInspect enables the visual inspection of astronomical objects. Once a table is loaded, AstroInspect automatically downloads and displays image cutouts for each object from the selected sources. Images can be retrieved from several surveys (detailed in Sec. \ref{sec:data}) and displayed simultaneously to facilitate comparative analyses, since the surveys can differ in characteristics, such as spatial or spectral coverage, number of filters, and observational depth.

AstroInspect can also display both local and remote images from user-defined sources. To enable this functionality, the user specifies a table column containing a resource identifier (RI) that uniquely defines the image location for each table row. Optional prefix and suffix strings may also be provided. AstroInspect constructs the image address for each object by concatenating the prefix, RI, and suffix in sequence. Any image format supported by modern web browsers can be loaded (e.g., JPG, PNG, GIF, WEBP).

For example, if the user specifies a RI column which contains values such as \texttt{J012502.90+002640.2}, and the user sets the prefix to \texttt{https://example.com/i/} and the suffix to \texttt{.jpg}, AstroInspect generates URLs of the form \texttt{https://example.com/i/J012502.90+002640.2.jpg} for each object. The same mechanism applies to local images, where the prefix corresponds to a directory path and the suffix specifies the file extension.

\subsection{Spectral energy distribution visualization}
\label{sec:sed}
Spectroscopic SEDs are retrieved from SDSS, DESI, or LAMOST. For the many objects lacking spectroscopy, photometric SEDs provide a crucial alternative. AstroInspect leverages the broad spectral coverage of S-PLUS, with its 12-band filter system, to generate these photometric SEDs in real time.

\subsection{Exploratory data analysis tool}
\label{sec:analysis}
The statistical visualization tools included in AstroInspect are scatter plots, histograms, and color-color diagrams. These graphs can be used to refine data samples by removing unreliable objects for visual inspection. Additionally, AstroInspect integrates the Aladin-Lite \citep{aladin-lite} package within the interface, thus enabling users to visualize the spatial arrangement of objects in the sky, overlay additional datasets, and interact with sky atlases.

\subsection{Interaction with other Virtual Observatory applications}
\label{sec:interop}
AstroInspect supports communication with other VO-compliant tools via Simple Application Messaging Protocol (SAMP; \citealp{samp, samp-ivoa}). SAMP enables real-time messaging between applications, allowing seamless data exchange without manual file handling. Through this protocol, AstroInspect can load tables from other graphical applications (e.g., Aladin or TOPCAT), or programmatically using packages such as Astropy, which provides Python-based interface to SAMP messaging.

\subsection{Object classification}
\label{sec:classification}
The classification tool in AstroInspect allows users to define custom categories and assign them to objects within the catalog, with the results exportable as a table. This functionality supports a broad range of scientific applications, such as morphological classification, catalog cleaning, and targeted sample selection. For example, Fig. \ref{fig:screen} illustrates a use case in which the ``class'' column includes the user-defined labels ``elliptical'' and ``spiral.'' To improve efficiency during large catalog inspections, users may also assign keyboard shortcuts to each class, reducing repetitive interactions.

\begin{figure*}[t!]
    \centering
    \includegraphics[width=\textwidth]{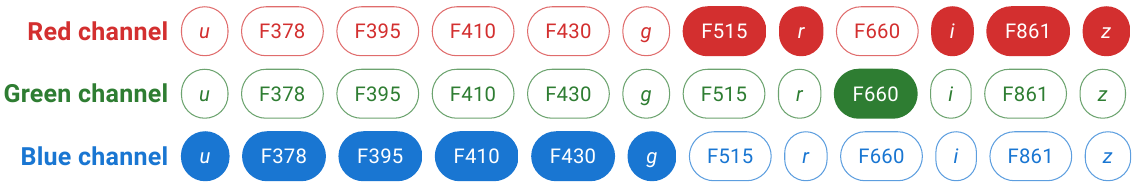}
    \caption{Screenshot of the AstroInspect interface that allows mapping each of the 12 filters of the S-PLUS photometric system ($u$, F378, F395, F410, F430, $g$, F515, $r$, F660, $i$, F861, $z$) to a channel in the RGB (red, green, and blue) color space. The configuration shows the use of filters $u$, F378, F395, F410, F430, and $g$ in the blue channel, F660 (H$\alpha$) alone in the green channel, and F515, $r$, $i$, F861, and $z$ in the red channel. The flexibility of this visualization tool allows users to create color composites that determine how physical information will be visually encoded in the image.}
    \label{fig:color-map}
\end{figure*}

\begin{figure*}[t!]
    \centering
    \includegraphics[width=\textwidth]{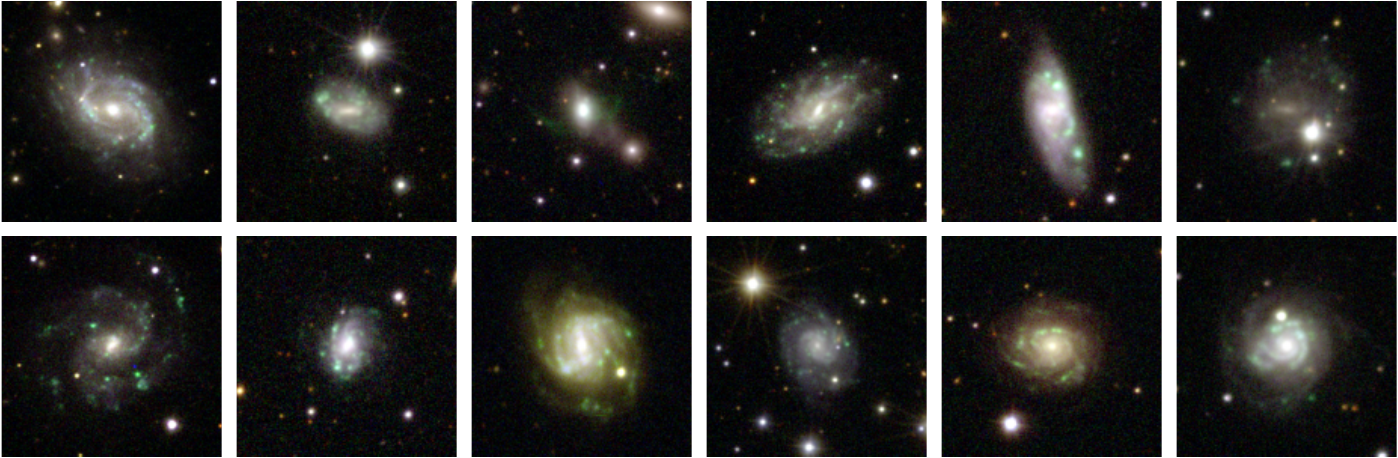}
    \caption{Mosaic of 12 H$\alpha$ emission galaxies in the direction of the Hydra I cluster found by visual inspection using AstroInspect. The images were created from the configuration indicated in Fig. \ref{fig:color-map}, where only the F660 filter (H$\alpha$) was mapped to the green channel of the RGB image. This visualization setup enables a rapid identification of H$\alpha$ emission regions.}
    \label{fig:halpha}
\end{figure*}

\section{Science case study: emission-line galaxies in the direction of the Hydra I cluster}
\label{sec:sci}
Galaxy clusters are among the most massive gravitationally bound structures in the universe, comprising hundreds to thousands of galaxies embedded within a hot, diffuse intracluster medium. Interactions between cluster galaxies and the intracluster medium -- including mechanisms such as ram-pressure stripping, strangulation, and tidal forces -- play a critical role in driving morphological transformations and quenching star formation in member galaxies \citep{gunn1972, gallagher1972, moore1996, jaffe2015}. In this context, visual inspection of galaxies in clusters remains a valuable approach for identifying and characterizing the wide range of morphological changes induced by environmental effects.

In this science example, we show the use of AstroInspect to build a catalog of galaxies with H$\alpha$ emission in the direction of the Hydra I cluster (also known as Abell 1060; \citealp{abell-clusters}) by visual inspection. Hydra I is of particular interest for this study, as its proximity enables the H$\alpha$ emission of disk galaxies in the cluster to be detected by the F660 filter and thus be very well highlighted in the 12-band RGB images from the S-PLUS survey, as shown in previous studies \citep{lima-dias2020,lima-dias2023}.

We adopt $\alpha,\delta$ = 10:36:41.8, -27:31:28 (J2000) as central position of Hydra I \citep{panagoulia2014}, a heliocentric redshift $z_\mathrm{cluster} = 0.0126$ \citep{struble1999}, and a virial radius $R_{200} \sim 1.44 \pm 0.08$ Mpc \citep{bohringer2002}. Considering that Hydra I is at a distance of 59 Mpc \citep{jorgensen1996}, we calculated the angular virial radius projecting \virial~on the sky at 59 Mpc, giving $R_{200} \sim 1.4$ deg. The value of $5R_{200}$ is 7.2 Mpc (7 deg). All magnitudes are in the AB system \citep{ab-system}.

\subsection{Data}
\label{sec:sci-data}
We used the catalogs from the 5$^{th}$ data release of S-PLUS (Lima et al., 2026, in preparation) to select a sample of galaxy candidates for H$\alpha$ emission, including the main photometric catalog and the Value Added Catalogs (VACs) of photometric redshift (photo-z) determinations \citep{lima2022} and galaxy-star-quasar classifications \citep{nakazono2021}. The sample was cross-matched with the southern hemisphere redshift compilation \citep{erik-redshifts} to include the spectroscopic redshift for those objects that have it. To inspect the selected objects, we used images from S-PLUS and LS, as well as photometric and spectroscopic SEDs from S-PLUS and SDSS, respectively.

\subsection{Method}
First, we select a sample of S-PLUS objects by making a cone centered at the central position of Hydra I with a radius equals to \virialf~and applying the following constraints:
\begin{itemize}
    \item \texttt{r\_auto} $<$ 21
    \item $0.001<$ \texttt{zml} $< 0.0326$
    \item \texttt{PROB\_GAL} $\geq p$
\end{itemize}

\texttt{r\_auto} is the $r$-band magnitude with auto aperture, \texttt{zml} is the photo-z, and \texttt{PROB\_GAL} is the probability to be a galaxy.

We considered 0.02 as the maximum difference between the Hydra I cluster's mean spectroscopic redshift and the object's photo-z (\texttt{zml}) that an object can be considered a candidate for cluster membership. We also imposed the minimum value of 0.001 to the photo-z range because we noted, through visual inspection, that all objects with photo-z below this value are stars.

The probability threshold $p$ varies depending on the galaxy's $r$-band magnitude. Specifically, it is set to: 0.90 for magnitudes brighter than 16; 0.64 for magnitudes between 16 and 17; 0.92 for 17 to 18; 0.72 for 18 to 19; 0.58 for 19 to 20; and 0.30 for magnitudes fainter than 20.

Applying these constraints resulted in a sample containing 3,787 objects. We inspected the $F660 - (r + i) / 2$ vs $F660$ color-magnitude diagram for this sample and made the following cut: 

\begin{itemize}
    \item $10 <$ \texttt{F660\_auto} $< 17$
    \item $-5 <$ \texttt{F660\_auto} $-$ (\texttt{r\_auto} $+$ \texttt{i\_auto}) $/\ 2 < 5$
\end{itemize}

That resulted in a subsample of 981 objects which we, finally, inspected using AstroInspect. \texttt{F660\_auto} and \texttt{i\_auto} are the F660 and i-band magnitude with auto aperture, respectively.

We configured the stamp production through the AstroInspect interface to highlight H$\alpha$-emission regions by mapping only the F660 filter to the green channel of the RGB image, as shown in Fig. \ref{fig:color-map}. This allows for a quick identification of regions of interest during visual inspection.

\subsection{Results}
The visual inspection of candidate galaxies yielded a final sample of 80 H$\alpha$ emission-line galaxies in the direction of the Hydra I within the whole circle of radius $5R_{200}$. Of these, 63 have spec-z derived from \citealp{erik-redshifts}, while the remaining 17 contains only photo-z derived from S-PLUS. Both the spec-z and photo-z measurements differ from $z_\mathrm{cluster}$ by no more than 0.02, supporting their association with the Hydra I cluster environment. A subset of these galaxies is presented in the mosaic in Fig. \ref{fig:halpha}, and the complete catalog is available in Table \ref{tab:emission}. These results provide resources for future studies.

\section{Discussion}
\label{sec:discussion}

\subsection{Case study analysis}
H$\alpha$ emission is a key tracer of star-forming regions, as it originates from recombination in H\textsc{ii} regions ionized by young, massive OB-type stars. Its presence indicates the existence of ionized gas and ongoing or star formation activity within galaxies. The high sensitivity of the S-PLUS F660 narrow-band filter enables the detection of H$\alpha$ emission from both extended star-forming regions in galactic disks and compact nuclear sources, making it a powerful tool for probing the spatial distribution and intensity of star formation across different galactic environments.

Visual selection of emission-line galaxies was essential to ensure the accurate identification of systems exhibiting significant H$\alpha$ emission, particularly given the morphological diversity and varying surface brightness of these features. As shown in Fig. \ref{fig:halpha}, the H$\alpha$ emission exhibits a wide range of morphologies, from patchy, clumpy structures in spiral arms and irregular galaxies to more centrally concentrated emission in galactic nuclei. This diversity likely reflects different stages of star formation, feedback processes, and evolutionary histories, as well as variations in galaxy mass and morphology. The observed patterns highlight the importance of visual inspection in complementing automated methods, allowing a more nuanced assessment of the physical conditions and evolutionary states of star-forming galaxies.

\begin{figure}[!t]
    \centering
    \includegraphics[width=\linewidth]{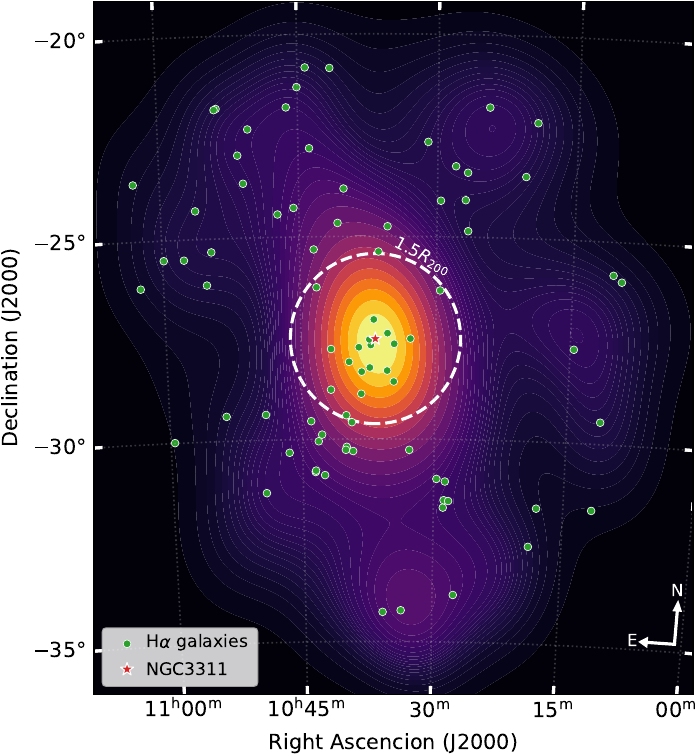}
    \caption{Comparison of the sample of 80 galaxies exhibiting H$\alpha$ emission (green dots) with the galaxy density in the Hydra I region. For the density visualization in the background, only galaxies with photo-z or spec-z between 0.001 and 0.0326 within a clustercentric distance of 5R$_{200}$ were included. The dashed white circle delimits the $1.5$\virial~radius, and the red star indicates the position of the NGC3311, the Hydra's brightest cluster galaxy (BCG).}
    \label{fig:hydra_density}
\end{figure}

Fig. \ref{fig:hydra_density} shows the spatial distribution of H$\alpha$-emitting galaxies, identified through visual inspection, overlaid on the density of galaxies in the Hydra I region. The density map shown in the background considers only galaxies with redshift (spectroscopic, or photometric for those objects without known spectroscopy) in the range between 0.001 and 0.0326, and at a maximum distance of \virialf~from the cluster center. 

Within $1.5\,R_{200}$, H$\alpha$ emitters are preferentially concentrated toward the southeast, while beyond this radius their distribution remains non-uniform. The resulting catalog provides a basis for future investigations of potential H$\alpha$-emitting substructures in Hydra I using complementary data. 

We noted that highlighting the H$\alpha$ component of galaxies directly in the AstroInspect interface greatly simplifies the identification of emission-line galaxies, enabling straightforward visual inspection.

Finally, we emphasize that this case study exemplifies the effectiveness of the  AstroInspect in supporting visual inspection tasks, specially in galaxy morphology research. Its integrated interface enables a comprehensive galaxy characterization, combining imaging with photometric and spectroscopic information. For a more detailed discussion of the Hydra I cluster, we refer the reader to the literature. \cite{lima-dias2020,lima-dias2023} investigate the physical and structural properties of Hydra I, including analyses of H$\alpha$-emitting galaxies. The forthcoming S-PLUS Clusters/Groups And their Large-scale Environments (SCALE; Mendes de Oliveira et al. 2026, in preparation) project will provide a comprehensive characterization of dozens of galaxy systems in the southern sky using combined spectroscopic and S-PLUS 12-band photometric data. With respect to Hydra I, this effort includes a detailed analysis of its dynamical properties (A. Ribeiro et al., in preparation) and a broader investigation of the large-scale environment of the Hydra--Antlia supercluster (A. Smith-Castelli et al., in preparation).

\subsection{Benefits and trade-offs}
\label{sec:benefits-tradeoffs}
Implementing AstroInspect as a web application has a direct impact on the user experience. The tool requires no compilation or installation, thereby enhancing accessibility. Its only prerequisite is a web browser, making it independent of operating systems and pre-installed system libraries. However, this design choice also constrains AstroInspect's functionality to the web technologies supported by browsers, as standardized by the World Wide Web Consortium (W3C).

Cross-Origin Resource Sharing (CORS) is a W3C specification that defines a browser-enforced security mechanism for controlling cross-origin HTTP requests. Under this model, a web service must explicitly allow its resources to be accessed by applications originating from other domains. Otherwise, the browser blocks the request, even though the same resources may be accessible to desktop applications or standalone scripts. As a result, AstroInspect can interact only with external services that explicitly permit cross-origin access. This limitation was considered during system design, but a web-based implementation was adopted to prioritize accessibility, portability, and ease of use. A technical description of the CORS mechanism is provided in Appendix \ref{ap:cors}.

Another limitation inherent to the use of web technologies is the limited access to the user's file system, which directly affects the implementation of performance-enhancing mechanisms such as long-term caching. Nevertheless, emerging web API specifications, such as the File System Access API\footnote{\url{https://wicg.github.io/file-system-access}}, may help mitigate this constraint once they achieve a broader adoption across browsers.

\subsection{Comparison with other tools}
TOPCAT is a well-established tool in astronomical research, providing robust functionalities for performing complex catalog operations. Despite its versatility, the visualization of image stamps in TOPCAT is limited to one object at a time, requiring repeated user input and reducing efficiency in tasks involving large samples.

Zooniverse\footnote{\url{https://zooniverse.org}} is a widely used platform for visual inspection in astronomy, with projects such as Galaxy Zoo \citep{gz,gz1} serving as notable examples. Its classification strategy is based on crowdsourcing, where contributions from many users are aggregated to produce reliable results. This approach has proven highly effective for large-scale citizen science initiatives \citep{tssc-platforms,tssc-natural}.

In contrast to TOPCAT and Zooniverse, AstroInspect integrates catalog data and information from external services within an unified interface, streamlining the inspection process. This approach allows for use cases not directly supported by those tools, such as the quick classification of objects in catalogs by a single user.

\subsection{Limitations}
\label{sec:limitations}
Tests conducted on various tables indicate that AstroInspect experiences performance degradation when handling tables exceeding 10 million cells. Additionally, AstroInspect relies on external services described in Section \ref{sec:data} and the unavailability of these services will directly impact the content displayed within the application.

\section{Conclusion and future work}
\label{sec:conclusion}
Research groups working on galaxy morphology, particularly those relying on visual inspection, often require custom tools tailored to their specific workflows. AstroInspect was initially developed to support internal studies within our group, and after being successfully employed in multiple projects, it has now matured into a general-purpose platform that can benefit the broader astronomical community.

In this work, we presented the first public release of the AstroInspect, a web-based system designed to streamline the visual inspection and scientific analysis of astronomical sources by integrating heterogeneous data products, such as images, spectra, and photometry, into a unified environment. Automated data retrieval, multi-survey integration, and configurable visualization options enhance efficiency in use cases such as morphological classification, catalog refinement, and exploratory analysis.

AstroInspect employs Pyodide, a WebAssembly-based Python runtime, to execute the scientific Python ecosystem directly in the browser. This client-side approach, increasingly adopted in scientific applications, provides access to widely used astronomical Python libraries without local installation or dedicated backend infrastructure. The system design presented in this work is adaptable to a range of use cases; for example, in data-intensive time-domain surveys such as the Vera C. Rubin Observatory's Legacy Survey of Space and Time (LSST; \citealp{lsst}), it can support the development of new applications designed for  rapid visual inspection, preliminary characterization, and follow-up prioritization of transient events.

We exemplified the scientific value of AstroInspect through a case study targeting H$\alpha$ emission-line galaxies in the direction of the Hydra I cluster. Beginning with a photometric catalog of 981 S-PLUS candidates, the AstroInspect's multi-band visualization and filtering capabilities enabled the rapid identification of 80 confirmed emission-line galaxies. This exercise highlights the effectiveness of the AstroInspect in supporting visual inspection tasks and provides a useful sample for studies of environmental effects in the Hydra I cluster.

Looking ahead, AstroInspect's modular architecture supports straightforward integration with data from next-generation facilities. Upcoming wide-field spectroscopic surveys, such as the William Herschel Telescope Enhanced Area Velocity Explorer (WEAVE; \citealp{weave}), the 4-metre Multi-Object Spectroscopic Telescope (4MOST; \citealp{4most}), and the Multi-Object Optical and Near-IR Spectrograph (MOONS; \citealp{moons}), will deliver extensive spectral datasets that can be directly incorporated to enhance visual inspection and analysis workflows.

\section*{Data availability}
The catalog of H$\alpha$ emission-line galaxies in the direction of the Hydra I cluster, described in Section \ref{sec:sci}, is available in the online version of this paper and the columns description are shown in Table \ref{tab:emission}. We also provide a digital version of the catalog in various formats on our page\footnote{\url{https://astroinspect.github.io/publications/hydra-halpha}}.

\begin{acknowledgments}
The S-PLUS project, including the T80-South robotic telescope and the S-PLUS scientific survey, was founded as a partnership between the Fundação de Amparo à Pesquisa do Estado de São Paulo (FAPESP), the Observatório Nacional (ON), the Federal University of Sergipe (UFS), and the Federal University of Santa Catarina (UFSC), with important financial and practical contributions from other collaborating institutes in Brazil, Chile (Universidad de La Serena), and Spain (Centro de Estudios de Física del Cosmos de Aragón, CEFCA). We further acknowledge financial support from the São Paulo Research Foundation (FAPESP), the Brazilian National Research Council (CNPq), the Coordination for the Improvement of Higher Education Personnel (CAPES), the Carlos Chagas Filho Rio de Janeiro State Research Foundation (FAPERJ), and the Brazilian Innovation Agency (FINEP).

The S-PLUS collaboration members are grateful for the contributions from CTIO staff in helping in the construction, commissioning, and maintenance of the T80-South telescope and camera. We are also indebted to Rene Laporte, INPE, and Keith Taylor for their essential contributions to the project. From CEFCA, we particularly would like to thank Antonio Marín-Franch for his invaluable contributions in the early phases of the project, David Cristóbal-Hornillos and his team for their help with the installation of the data reduction package JYPE version 0.9.9, César Íñiguez for providing 2D measurements of the filter transmissions, and all other staff members for their support with various aspects of the project.

Funding for the Sloan Digital Sky Survey V has been provided by the Alfred P. Sloan Foundation, the Heising-Simons Foundation, the National Science Foundation, and the Participating Institutions. SDSS acknowledges support and resources from the Center for High-Performance Computing at the University of Utah. SDSS telescopes are located at Apache Point Observatory, funded by the Astrophysical Research Consortium and operated by New Mexico State University, and at Las Campanas Observatory, operated by the Carnegie Institution for Science.

The Legacy Surveys consist of three individual and complementary projects: the Dark Energy Camera Legacy Survey (DECaLS; Proposal ID \#2014B-0404; PIs: David Schlegel and Arjun Dey), the Beijing-Arizona Sky Survey (BASS; NOAO Prop. ID \#2015A-0801; PIs: Zhou Xu and Xiaohui Fan), and the Mayall z-band Legacy Survey (MzLS; Prop. ID \#2016A-0453; PI: Arjun Dey). DECaLS, BASS and MzLS together include data obtained, respectively, at the Blanco telescope, Cerro Tololo Inter-American Observatory, NSF’s NOIRLab; the Bok telescope, Steward Observatory, University of Arizona; and the Mayall telescope, Kitt Peak National Observatory, NOIRLab. Pipeline processing and analyses of the data were supported by NOIRLab and the Lawrence Berkeley National Laboratory (LBNL). The Legacy Surveys project is honored to be permitted to conduct astronomical research on Iolkam Du'ag (Kitt Peak), a mountain with particular significance to the Tohono O’odham Nation.

DESI construction and operations is managed by the Lawrence Berkeley National Laboratory. This research is supported by the U.S. Department of Energy, Office of Science, Office of High-Energy Physics, under Contract No. DE–AC02–05CH11231, and by the National Energy Research Scientific Computing Center, a DOE Office of Science User Facility under the same contract. Additional support for DESI is provided by the U.S. National Science Foundation, Division of Astronomical Sciences under Contract No. AST-0950945 to the NSF’s National Optical-Infrared Astronomy Research Laboratory; the Science and Technology Facilities Council of the United Kingdom; the Gordon and Betty Moore Foundation; the Heising-Simons Foundation; the French Alternative Energies and Atomic Energy Commission (CEA); the National Council of Science and Technology of Mexico (CONACYT); the Ministry of Science and Innovation of Spain, and by the DESI Member Institutions. The DESI collaboration is honored to be permitted to conduct astronomical research on Iolkam Du’ag (Kitt Peak), a mountain with particular significance to the Tohono O’odham Nation.

Guoshoujing Telescope (the Large Sky Area Multi-Object Fiber Spectroscopic Telescope LAMOST) is a National Major Scientific Project built by the Chinese Academy of Sciences. Funding for the project has been provided by the National Development and Reform Commission. LAMOST is operated and managed by the National Astronomical Observatories, Chinese Academy of Sciences.

This research has made use of the hips2fits, a tool developed at CDS, Strasbourg, France aiming at extracting FITS images from HiPS sky maps with respect to a WCS.

This research uses services or data provided by the SPectra Analysis and Retrievable Catalog Lab (SPARCL) and the Astro Data Lab, which are both part of the Community Science and Data Center (CSDC) program at NSF National Optical-Infrared Astronomy Research Laboratory. NOIRLab is operated by the Association of Universities for Research in Astronomy (AURA), Inc. under a cooperative agreement with the National Science Foundation.

N.M.C. thanks the Coordenação de Aperfeiçoamento de Pessoal de Nível Superior -- Brasil (CAPES) -- Finance Code 88887.133104/2025-00. C.M.d.O. thanks Pesquisa do Estado de São Paulo (FAPESP) for funding through grants 2019/26492-3 and 2023/05087-9 and Conselho Nacional de Desenvolvimento Científico e Tecnológico (CNPq) through grant 309209/2019-6. A.C.K. thanks FAPESP for the support grant 2024/05467-9 and the CNPq. A.V.S.C. thanks financial support from Consejo Nacional de Investigaciones Científicas y Técnicas (CONICET) (PIP 1504), Agencia I+D+i (PICT 2019–03299) and Universidad Nacional de La Plata (Argentina). M.S.C acknowledges funding from FAPESP grant 2025/12629-8. R.D. thanks support by the ANID BASAL project FB210003.
\end{acknowledgments}

\appendix
\section{The emission-line catalog}
\label{ap:catalog}
We release a catalog of 80 H$\alpha$ emission-line galaxies in the direction of the Hydra I cluster, selected according to the procedure described in Section \ref{sec:sci}. The catalog includes multiwavelength photometry from S-PLUS DR5 and LS DR10, as well as spectroscopic redshifts when available. A description of the catalog columns is provided in Table \ref{tab:emission}.

\startlongtable
\begin{deluxetable*}{cclp{0.6\linewidth}}
\digitalasset
\tablewidth{\linewidth}
\tablecaption{Column descriptions of the catalog containing the H$\alpha$ emission-line galaxies in the direction of the Hydra I cluster selected through visual inspection\label{tab:emission}}
\tablehead{
   \colhead{\#} & \colhead{Unit} & \colhead{Column name} & \colhead{Description}
}
\startdata
1 &  & id & Catalog unique identifier \\
2 & deg & ra & Right Ascension in decimal degrees (J2000) \\
3 & deg & dec & Declination in decimal degrees (J2000) \\
4 & mag & sp\_mag\_u & S-PLUS DR5 magnitude in the $u$ band\\
5 & mag & sp\_mag\_g & S-PLUS DR5 magnitude in the $g$ band\\
6 & mag & sp\_mag\_r & S-PLUS DR5 magnitude in the $r$ band\\
7 & mag & sp\_mag\_i & S-PLUS DR5 magnitude in the $i$ band\\
8 & mag & sp\_mag\_z & S-PLUS DR5 magnitude in the $z$ band\\
9 & mag & sp\_mag\_F378 & S-PLUS DR5 magnitude in the F378 band (Balmer jump / [O\textsc{ii}])\\
10 & mag & sp\_mag\_F395 & S-PLUS DR5 magnitude in the F395 band (Ca H + K)\\
11 & mag & sp\_mag\_F410 & S-PLUS DR5 magnitude in the F410 band (H$\delta$)\\
12 & mag & sp\_mag\_F430 & S-PLUS DR5 magnitude in the F430 band (G band)\\
13 & mag & sp\_mag\_F515 & S-PLUS DR5 magnitude in the F515 band (Mg b triplet)\\
14 & mag & sp\_mag\_F660 & S-PLUS DR5 magnitude in the F660 band (H$\alpha$)\\
15 & mag & sp\_mag\_F861 & S-PLUS DR5 magnitude in the F861 band (Ca triplet)\\
16--27 & mag & sp\_mag\_err\_\texttt{[band]} & S-PLUS DR5 magnitude error, where \texttt{[band]} is one of: $u$, $g$, $r$, $i$, $z$, F378, F395, F410, F430, F515, F660, F861\\
28 & deg & sp\_A & S-PLUS DR5 profile RMS along the major axis\\
29 & deg & sp\_B & S-PLUS DR5 profile RMS along the minor axis\\
30 & deg & sp\_PA & S-PLUS DR5 position angle (CCW/World-x)\\
31 & deg & sp\_radius\_petro & S-PLUS DR5 petrosian apertures\\
32 & deg & sp\_radius\_50 & S-PLUS DR5 radius enclosing 50\%  of the total flux\\
33 & deg & sp\_radius\_90 & S-PLUS DR5 radius enclosing 90\%  of the total flux\\
34 &  & sp\_photoz & S-PLUS DR5 photometric redshift\\
35 &  & sp\_photoz\_odds & S-PLUS DR5 photometric redshift odds\\
36 & mag & ls\_mag\_g & LS DR10 magnitude in the $g$ band \\
37 & mag & ls\_mag\_r & LS DR10 magnitude in the $r$ band \\
38 & mag & ls\_mag\_i & LS DR10 magnitude in the $i$ band \\
39 & mag & ls\_mag\_z & LS DR10 magnitude in the $z$ band \\
40 &  & ls\_type & LS DR10 morphological type (for details, see \url{https://www.legacysurvey.org/dr10/description})\\
41 &  & lit\_redshift & Spectroscopic redshift\\
42 &  & lit\_redshift\_err & Spectroscopic redshift error\\
43 &  & lit\_source & Spectroscopic redshift source from literature\\
44 & km/s & velocity & Velocity derived from spectroscopic redshift\\
45 & km/s & velocity\_err & Velocity error\\
46 & km/s & velocity\_offset & Velocity difference between the object and the Hydra I, only available for objects with known spectroscopic redshift\\
47 & deg & dist\_proj & Sky projected angular distance between the object and the Hydra I cluster center\\
48 & Mpc & dist & Distance between the object and the Hydra I cluster center, only available for objects with known spectroscopic redshift\\
\enddata
\tablecomments{Columns prefixed with \texttt{sp} are drawn from S-PLUS DR5 (E. V. Lima et al. 2026, in preparation), while those prefixed with \texttt{ls} originate from Legacy Survey DR10 \citep{legacy}. Columns prefixed with \texttt{lit} originate from redshift compilation from the literature \citep{erik-redshifts}. Columns without these prefixes were derived in this work. All magnitudes are reported in the AB system, and the S-SPLUS magnitudes included in the catalog are only the auto-aperture values. This table is published in its entirety in the electronic edition of the {\it Astronomical Journal}. The columns description is shown here for guidance regarding its form and content.}
\end{deluxetable*}

\section{Technical description of the cross-origin resource sharing}
\label{ap:cors}
Cross-Origin Resource Sharing (CORS) is a browser-enforced security model configured through server-side HTTP response headers. For cross-origin requests, the browser determines whether a web application may access the response based solely on these headers. If the server does not explicitly authorize the requesting origin, the browser blocks the response from the application, even if the server successfully processed the request.

To illustrate the CORS mechanism in practice, we analyze a real HTTP response from the SkyServer \citep{skyserver,sciserver} service. Listing \ref{lst:http} presents the complete set of response headers returned by an HTTP request to SkyServer. The CORS-related headers appear in lines 10 to 12 of Listing \ref{lst:http}.

The \texttt{Access-Control-Allow-Origin} header specifies which origins are permitted to access the response from a web application. In this example, its value is set to the wildcard character \texttt{*}, indicating that the resource is accessible from any origin. The browser interprets this as permission to expose the response to all client-side web applications, regardless of their domain. If this header were absent or restricted to a different origin, the browser would block access to the response.

The \texttt{Access-Control-Allow-Headers} header defines which custom request headers are allowed in cross-origin requests. Here, the value lists \texttt{Content-Type}, \texttt{X-Auth-Token}, and \texttt{Accept}, indicating that requests including these headers are permitted. The \texttt{Access-Control-Allow-Methods} header specifies the HTTP methods (GET, POST, PUT, etc) allowed for cross-origin access. Its value is also set to \texttt{*}, meaning that all HTTP methods are accepted.

Together, these headers indicate that SkyServer explicitly allows cross-origin access without restrictions on origin or HTTP method. This example emphasizes that CORS does not constrain server-to-server communication. Instead, it governs how browsers expose cross-origin responses to client-side applications. Adopting permissive and well-defined CORS headers in astronomical web services is therefore a recommended best practice, as it enables browser-based scientific applications to access these services directly, broadening the ecosystem of web-enabled tools for astronomical data analysis.

\vspace{6pt}
\begin{lstlisting}[label=lst:http,
language=http, 
numbers=left, 
caption={HTTP response headers returned by the SkyServer service, highlighting the CORS-related fields in lines 10 to 12 that permit access to its resources by web-based applications. To produce this example, we used the \texttt{cURL} command-line HTTP client with the following call: \texttt{curl -IX GET "https://skyserver.sdss.org/dr19/SkyServerWS/SpectroQuery/ConeSpectro?\allowbreak~ra=344.5894\&dec=-1.1839\&radius=0.0833\&limit=1\&format=xml\&specparams=specObjID"}}]
HTTP/1.1 200 OK
cache-control: no-cache
pragma: no-cache
transfer-encoding: chunked
content-type: application/xml
expires: -1
server: Microsoft-IIS/10.0
x-aspnet-version: 4.0.30319
x-powered-by: ASP.NET
access-control-allow-origin: *
access-control-allow-headers: Content-Type, X-Auth-Token, Accept
access-control-allow-methods: *
date: Fri, 23 Jan 2026 11:59:22 GMT
strict-transport-security: max-age=63072000
content-security-policy: upgrade-insecure-requests
set-cookie: SERVERID=idieswww02; path=/
\end{lstlisting}





\bibliography{sample701}{}

@article{sdss,
title = "The Sloan Digital Sky Survey: Technical summary",
author = "York, {Donald G.} and J. Adelman and Anderson, {John E.} and Anderson, {Scott F.} and James Annis and Bahcall, {Neta A.} and Bakken, {J. A.} and Robert Barkhouser and Steven Bastian and Eileen Berman and Boroski, {William N.} and Steve Bracker and Charlie Briegel and Briggs, {John W.} and J. Brinkmann and Robert Brunner and Scott Burles and Larry Carey and Carr, {Michael A.} and Castander, {Francisco J.} and Bing Chen and Colestock, {Patrick L.} and Connolly, {A. J.} and Crocker, {J. H.} and Istv{\'a}n Csabai and Czarapata, {Paul C.} and Davis, {John Eric} and Mamoru Doi and Tom Dombeck and Daniel Eisenstein and Nancy Ellman and Elms, {Brian R.} and Evans, {Michael L.} and Xiaohui Fan and Federwitz, {Glenn R.} and Larry Fiscelli and Scott Friedman and Frieman, {Joshua A.} and Masataka Fukugita and Bruce Gillespie and Gunn, {James E.} and Gurbani, {Vijay K.} and {De Haas}, Ernst and Merle Haldeman and Harris, {Frederick H.} and J. Hayes and Heckman, {Timothy M.} and Hennessy, {G. S.} and Hindsley, {Robert B.} and Scott Holm and Holmgren, {Donald J.} and Huang, {Chi Hao} and Charles Hull and Don Husby and Ichikawa, {Shin Ichi} and Takashi Ichikawa and Z{\v e}ljko Ivezi{\'c} and Stephen Kent and Kim, {Rita S.J.} and E. Kinney and Mark Klaene and Kleinman, {A. N.} and S. Kleinman and Knapp, {G. R.} and John Korienek and Kron, {Richard G.} and Kunszt, {Peter Z.} and Lamb, {D. Q.} and B. Lee and Leger, {R. French} and Siriluk Limmongkol and Carl Lindenmeyer and Long, {Daniel C.} and Craig Loomis and Jon Loveday and Rich Lucinio and Lupton, {Robert H.} and Bryan Mackinnon and Mannery, {Edward J.} and Mantsch, {P. M.} and Bruce Margon and Peregrine Mcgehee and Mckay, {Timothy A.} and Avery Meiksin and Aronne Merelli and Monet, {David G.} and Munn, {Jeffrey A.} and Narayanan, {Vijay K.} and Thomas Nash and Eric Neilsen and Rich Neswold and Newberg, {Heidi Jo} and Nichol, {R. C.} and Tom Nicinski and Mario Nonino and Norio Okada and Sadanori Okamura and Ostriker, {Jeremiah P.} and Russell Owen and Pauls, {A. George} and John Peoples and Peterson, {R. L.} and Donald Petravick and Pier, {Jeffrey R.} and Adrian Pope and Ruth Pordes and Angela Prosapio and Ron Rechenmacher and Quinn, {Thomas R.} and Richards, {Gordon T.} and Richmond, {Michael W.} and Rivetta, {Claudio H.} and Rockosi, {Constance M.} and Kurt Ruthmansdorfer and Dale Sandford and Schlegel, {David J.} and Schneider, {Donald P.} and Maki Sekiguchi and Gary Sergey and Kazuhiro Shimasaku and Siegmund, {Walter A.} and Stephen Smee and Smith, {J. Allyn} and S. Snedden and R. Stone and Chris Stoughton and Strauss, {Michael A.} and Christopher Stubbs and Mark Subbarao and Szalay, {Alexander S.} and Istvan Szapudi and Szokoly, {Gyula P.} and Thakar, {Anirudda R.} and Christy Tremonti and Tucker, {Douglas L.} and Alan Uomoto and {Vanden Berk}, Dan and Vogeley, {Michael S.} and Patrick Waddell and Wang, {Shu I.} and Masaru Watanabe and Weinberg, {David H.} and Brian Yanny and Naoki Yasuda",
year = "2000",
month = sep,
doi = "10.1086/301513",
language = "English (US)",
volume = "120",
pages = "1579--1587",
journal = "Astronomical Journal",
issn = "0004-6256",
publisher = "IOP Publishing Ltd.",
number = "3",
}

@article{sdss-spec, 
title={THE MULTI-OBJECT, FIBER-FED SPECTROGRAPHS FOR THE SLOAN DIGITAL SKY SURVEY AND THE BARYON OSCILLATION SPECTROSCOPIC SURVEY}, 
volume={146}, 
DOI={10.1088/0004-6256/146/2/32}, 
number={2}, 
journal={The Astronomical Journal}, 
publisher={American Astronomical Society}, 
author={Smee,  Stephen A. and Gunn,  James E. and Uomoto,  Alan and Roe,  Natalie and Schlegel,  David and Rockosi,  Constance M. and Carr,  Michael A. and Leger,  French and Dawson,  Kyle S. and Olmstead,  Matthew D. and Brinkmann,  Jon and Owen,  Russell and Barkhouser,  Robert H. and Honscheid,  Klaus and Harding,  Paul and Long,  Dan and Lupton,  Robert H. and Loomis,  Craig and Anderson,  Lauren and Annis,  James and Bernardi,  Mariangela and Bhardwaj,  Vaishali and Bizyaev,  Dmitry and Bolton,  Adam S. and Brewington,  Howard and Briggs,  John W. and Burles,  Scott and Burns,  James G. and Castander,  Francisco Javier and Connolly,  Andrew and Davenport,  James R. A. and Ebelke,  Garrett and Epps,  Harland and Feldman,  Paul D. and Friedman,  Scott D. and Frieman,  Joshua and Heckman,  Timothy and Hull,  Charles L. and Knapp,  Gillian R. and Lawrence,  David M. and Loveday,  Jon and Mannery,  Edward J. and Malanushenko,  Elena and Malanushenko,  Viktor and Merrelli,  Aronne James and Muna,  Demitri and Newman,  Peter R. and Nichol,  Robert C. and Oravetz,  Daniel and Pan,  Kaike and Pope,  Adrian C. and Ricketts,  Paul G. and Shelden,  Alaina and Sandford,  Dale and Siegmund,  Walter and Simmons,  Audrey and Smith,  D. Shane and Snedden,  Stephanie and Schneider,  Donald P. and SubbaRao,  Mark and Tremonti,  Christy and Waddell,  Patrick and York,  Donald G.}, 
year={2013}, 
month=jul, 
pages={32} 
}

@article{sdss-filters, 
title={The Sloan Digital Sky Survey Photometric System}, 
volume={111}, 
DOI={10.1086/117915}, 
journal={The Astronomical Journal}, 
publisher={American Astronomical Society}, 
author={Fukugita, M. and Ichikawa, T. and Gunn, J. E. and Doi, M. and Shimasaku, K. and Schneider, D. P.}, 
year={1996}, 
month=apr, 
pages={1748} 
}

@article{sdss-telescope,
  title = {The 2.5 m Telescope of the Sloan Digital Sky Survey},
  volume = {131},
  ISSN = {1538-3881},
  DOI = {10.1086/500975},
  number = {4},
  journal = {The Astronomical Journal},
  publisher = {American Astronomical Society},
  author = {Gunn,  James E. and Siegmund,  Walter A. and Mannery,  Edward J. and Owen,  Russell E. and Hull,  Charles L. and Leger,  R. French and Carey,  Larry N. and Knapp,  Gillian R. and York,  Donald G. and Boroski,  William N. and Kent,  Stephen M. and Lupton,  Robert H. and Rockosi,  Constance M. and Evans,  Michael L. and Waddell,  Patrick and Anderson,  John E. and Annis,  James and Barentine,  John C. and Bartoszek,  Larry M. and Bastian,  Steven and Bracker,  Stephen B. and Brewington,  Howard J. and Briegel,  Charles I. and Brinkmann,  Jon and Brown,  Yorke J. and Carr,  Michael A. and Czarapata,  Paul C. and Drennan,  Craig C. and Dombeck,  Thomas and Federwitz,  Glenn R. and Gillespie,  Bruce A. and Gonzales,  Carlos and Hansen,  Sten U. and Harvanek,  Michael and Hayes,  Jeffrey and Jordan,  Wendell and Kinney,  Ellyne and Klaene,  Mark and Kleinman,  S. J. and Kron,  Richard G. and Kresinski,  Jurek and Lee,  Glenn and Limmongkol,  Siriluk and Lindenmeyer,  Carl W. and Long,  Daniel C. and Loomis,  Craig L. and McGehee,  Peregrine M. and Mantsch,  Paul M. and Neilsen,  Jr.,  Eric H. and Neswold,  Richard M. and Newman,  Peter R. and Nitta,  Atsuko and Peoples,  Jr.,  John and Pier,  Jeffrey R. and Prieto,  Peter S. and Prosapio,  Angela and Rivetta,  Claudio and Schneider,  Donald P. and Snedden,  Stephanie and Wang,  Shu-i},
  year = {2006},
  month = apr,
  pages = {2332–2359}
}

@article{legacy,
  title = {Overview of the DESI Legacy Imaging Surveys},
  volume = {157},
  ISSN = {1538-3881},
  DOI = {10.3847/1538-3881/ab089d},
  number = {5},
  journal = {The Astronomical Journal},
  publisher = {American Astronomical Society},
  author = {Dey,  Arjun and Schlegel,  David J. and Lang,  Dustin and Blum,  Robert and Burleigh,  Kaylan and Fan,  Xiaohui and Findlay,  Joseph R. and Finkbeiner,  Doug and Herrera,  David and Juneau,  Stéphanie and Landriau,  Martin and Levi,  Michael and McGreer,  Ian and Meisner,  Aaron and Myers,  Adam D. and Moustakas,  John and Nugent,  Peter and Patej,  Anna and Schlafly,  Edward F. and Walker,  Alistair R. and Valdes,  Francisco and Weaver,  Benjamin A. and Yèche,  Christophe and Zou,  Hu and Zhou,  Xu and Abareshi,  Behzad and Abbott,  T. M. C. and Abolfathi,  Bela and Aguilera,  C. and Alam,  Shadab and Allen,  Lori and Alvarez,  A. and Annis,  James and Ansarinejad,  Behzad and Aubert,  Marie and Beechert,  Jacqueline and Bell,  Eric F. and BenZvi,  Segev Y. and Beutler,  Florian and Bielby,  Richard M. and Bolton,  Adam S. and Briceño,  César and Buckley-Geer,  Elizabeth J. and Butler,  Karen and Calamida,  Annalisa and Carlberg,  Raymond G. and Carter,  Paul and Casas,  Ricard and Castander,  Francisco J. and Choi,  Yumi and Comparat,  Johan and Cukanovaite,  Elena and Delubac,  Timothée and DeVries,  Kaitlin and Dey,  Sharmila and Dhungana,  Govinda and Dickinson,  Mark and Ding,  Zhejie and Donaldson,  John B. and Duan,  Yutong and Duckworth,  Christopher J. and Eftekharzadeh,  Sarah and Eisenstein,  Daniel J. and Etourneau,  Thomas and Fagrelius,  Parker A. and Farihi,  Jay and Fitzpatrick,  Mike and Font-Ribera,  Andreu and Fulmer,  Leah and G\"{a}nsicke,  Boris T. and Gaztanaga,  Enrique and George,  Koshy and Gerdes,  David W. and A Gontcho,  Satya Gontcho and Gorgoni,  Claudio and Green,  Gregory and Guy,  Julien and Harmer,  Diane and Hernandez,  M. and Honscheid,  Klaus and Huang,  Lijuan (Wendy) and James,  David J. and Jannuzi,  Buell T. and Jiang,  Linhua and Joyce,  Richard and Karcher,  Armin and Karkar,  Sonia and Kehoe,  Robert and Kneib,  Jean-Paul and Kueter-Young,  Andrea and Lan,  Ting-Wen and Lauer,  Tod R. and Guillou,  Laurent Le and Van Suu,  Auguste Le and Lee,  Jae Hyeon and Lesser,  Michael and Levasseur,  Laurence Perreault and Li,  Ting S. and Mann,  Justin L. and Marshall,  Robert and Martínez-Vázquez,  C. E. and Martini,  Paul and du Mas des Bourboux,  Hélion and McManus,  Sean and Meier,  Tobias Gabriel and Ménard,  Brice and Metcalfe,  Nigel and Muñoz-Gutiérrez,  Andrea and Najita,  Joan and Napier,  Kevin and Narayan,  Gautham and Newman,  Jeffrey A. and Nie,  Jundan and Nord,  Brian and Norman,  Dara J. and Olsen,  Knut A. G. and Paat,  Anthony and Palanque-Delabrouille,  Nathalie and Peng,  Xiyan and Poppett,  Claire L. and Poremba,  Megan R. and Prakash,  Abhishek and Rabinowitz,  David and Raichoor,  Anand and Rezaie,  Mehdi and Robertson,  A. N. and Roe,  Natalie A. and Ross,  Ashley J. and Ross,  Nicholas P. and Rudnick,  Gregory and Gaines,  Sasha and Saha,  Abhijit and Sánchez,  F. Javier and Savary,  Elodie and Schweiker,  Heidi and Scott,  Adam and Seo,  Hee-Jong and Shan,  Huanyuan and Silva,  David R. and Slepian,  Zachary and Soto,  Christian and Sprayberry,  David and Staten,  Ryan and Stillman,  Coley M. and Stupak,  Robert J. and Summers,  David L. and Tie,  Suk Sien and Tirado,  H. and Vargas-Magaña,  Mariana and Vivas,  A. Katherina and Wechsler,  Risa H. and Williams,  Doug and Yang,  Jinyi and Yang,  Qian and Yapici,  Tolga and Zaritsky,  Dennis and Zenteno,  A. and Zhang,  Kai and Zhang,  Tianmeng and Zhou,  Rongpu and Zhou,  Zhimin},
  year = {2019},
  month = apr,
  pages = {168}
}

@article{bass,
  title = {Project Overview of the Beijing–Arizona Sky Survey},
  volume = {129},
  ISSN = {1538-3873},
  DOI = {10.1088/1538-3873/aa65ba},
  number = {976},
  journal = {Publications of the Astronomical Society of the Pacific},
  publisher = {IOP Publishing},
  author = {Zou,  Hu and Zhou,  Xu and Fan,  Xiaohui and Zhang,  Tianmeng and Zhou,  Zhimin and Nie,  Jundan and Peng,  Xiyan and McGreer,  Ian and Jiang,  Linhua and Dey,  Arjun and Fan,  Dongwei and He,  Boliang and Jiang,  Zhaoji and Lang,  Dustin and Lesser,  Michael and Ma,  Jun and Mao,  Shude and Schlegel,  David and Wang,  Jiali},
  year = {2017},
  month = apr,
  pages = {064101}
}

@Article{topcat,
AUTHOR = {Taylor, Mark},
TITLE = {TOPCAT: Desktop Exploration of Tabular Data for Astronomy and Beyond},
JOURNAL = {Informatics},
VOLUME = {4},
YEAR = {2017},
NUMBER = {3},
ARTICLE-NUMBER = {18},
ISSN = {2227-9709},
DOI = {10.3390/informatics4030018}
}

@INPROCEEDINGS{saods9,
       author = {{Joye}, W.~A. and {Mandel}, E.},
        title = "{New Features of SAOImage DS9}",
    booktitle = {Astronomical Data Analysis Software and Systems XII},
         year = 2003,
       editor = {{Payne}, H.~E. and {Jedrzejewski}, R.~I. and {Hook}, R.~N.},
       series = {Astronomical Society of the Pacific Conference Series},
       volume = {295},
        month = jan,
        pages = {489},
       adsurl = {https://ui.adsabs.harvard.edu/abs/2003ASPC..295..489J}
}

@article{js9,
	author = {Matilsky, Terry},
	title = {{JS9: An interactive tool for teaching astrophysics}},
	journal = {Phys. Teach.},
	volume = {58},
	number = {8},
	pages = {602--603},
	year = {2020},
	month = nov,
	issn = {0031-921X},
	publisher = {AIP Publishing},
	doi = {10.1119/10.0002391}
}

@INPROCEEDINGS{aladin,
       author = {{Boch}, T. and {Oberto}, A. and {Fernique}, P. and {Bonnarel}, F.},
        title = "{Aladin: An Open Source All-Sky Browser}",
    booktitle = {Astronomical Data Analysis Software and Systems XX},
         year = 2011,
       editor = {{Evans}, I.~N. and {Accomazzi}, A. and {Mink}, D.~J. and {Rots}, A.~H.},
       series = {Astronomical Society of the Pacific Conference Series},
       volume = {442},
        month = jul,
        pages = {683},
       adsurl = {https://ui.adsabs.harvard.edu/abs/2011ASPC..442..683B}
}

@article{aladin-2,
   author = {Ph. Paillou and F. Bonnarel and F. Ochsenbein and M. Crézé},
   doi = {10.1017/S0074180900047628},
   issn = {0074-1809},
   journal = {Symposium - International Astronomical Union},
   pages = {347-351},
   publisher = {Cambridge University Press},
   title = {Aladin: An Interactive Deep Sky Mapping Facility},
   volume = {161},
   year = {1994},
}

@inproceedings{aladin-lite,
  TITLE = {{Aladin Lite v3: Behind the Scenes of a Major Overhaul}},
  AUTHOR = {Baumann, Matthieu and Boch, Thomas and Pineau, Fran{\c c}ois-Xavier and Fernique, Pierre and Bot, Caroline and Allen, Mark},
  URL = {https://hal.science/hal-03842974},
  BOOKTITLE = {{Astronomical Data Analysis Software and Systems XXX. ASP Conference Series}},
  ADDRESS = {virtual conference, Spain},
  EDITOR = {Jose Enrique Ruiz and Francesco Pierfedereci and Peter Teuben},
  SERIES = {Astronomical Data Analysis Software and Systems XXX. ASP Conference Series, Vol. 532, Proceedings of a virtual conference held 8-12 November 2020.},
  VOLUME = {532},
  PAGES = {7},
  YEAR = {2020},
  MONTH = Nov,
  HAL_ID = {hal-03842974},
  HAL_VERSION = {v1},
}

@article{hips,
  title = {Hierarchical progressive surveys: Multi-resolution HEALPix data structures for astronomical images,  catalogues,  and 3-dimensional data cubes},
  volume = {578},
  ISSN = {1432-0746},
  DOI = {10.1051/0004-6361/201526075},
  journal = {Astronomy \& Astrophysics},
  publisher = {EDP Sciences},
  author = {Fernique,  P. and Allen,  M. G. and Boch,  T. and Oberto,  A. and Pineau,  F-X. and Durand,  D. and Bot,  C. and Cambrésy,  L. and Derriere,  S. and Genova,  F. and Bonnarel,  F.},
  year = {2015},
  month = jun,
  pages = {A114}
}

@article{splus,
    author = {Mendes de Oliveira, C and Ribeiro, T and Schoenell, W and Kanaan, A and Overzier, R A and Molino, A and Sampedro, L and Coelho, P and Barbosa, C E and Cortesi, A and Costa-Duarte, M V and Herpich, F R and Hernandez-Jimenez, J A and Placco, V M and Xavier, H S and Abramo, L R and Saito, R K and Chies-Santos, A L and Ederoclite, A and Lopes de Oliveira, R and Gonçalves, D R and Akras, S and Almeida, L A and Almeida-Fernandes, F and Beers, T C and Bonatto, C and Bonoli, S and Cypriano, E S and Vinicius-Lima, E and de Souza, R S and Fabiano de Souza, G and Ferrari, F and Gonçalves, T S and Gonzalez, A H and Gutiérrez-Soto, L A and Hartmann, E A and Jaffe, Y and Kerber, L O and Lima-Dias, C and Lopes, P A A and Menendez-Delmestre, K and Nakazono, L M I and Novais, P M and Ortega-Minakata, R A and Pereira, E S and Perottoni, H D and Queiroz, C and Reis, R R R and Santos, W A and Santos-Silva, T and Santucci, R M and Barbosa, C L and Siffert, Beatriz B and Sodré, L, Jr and Torres-Flores, S and Westera, P and Whitten, D D and Alcaniz, J S and Alonso-García, Javier and Alencar, S and Alvarez-Candal, A and Amram, P and Azanha, L and Barbá, R H and Bernardinelli, P H and Borges Fernandes, M and Branco, V and Brito-Silva, D and Buzzo, M L and Caffer, J and Campillay, A and Cano, Z and Carvano, J M and Castejon, M and Cid Fernandes, R and Dantas, M L L and Daflon, S and Damke, G and de la Reza, R and de Melo de Azevedo, L J and De Paula, D F and Diem, K G and Donnerstein, R and Dors, O L and Dupke, R and Eikenberry, S and Escudero, Carlos G and Faifer, Favio R and Farías, H and Fernandes, B and Fernandes, C and Fontes, S and Galarza, A and Hirata, N S T and Katena, L and Gregorio-Hetem, J and Hernández-Fernández, J D and Izzo, L and Jaque Arancibia, M and Jatenco-Pereira, V and Jiménez-Teja, Y and Kann, D A and Krabbe, A C and Labayru, C and Lazzaro, D and Lima Neto, G B and Lopes, Amanda R and Magalhães, R and Makler, M and de Menezes, R and Miralda-Escudé, J and Monteiro-Oliveira, R and Montero-Dorta, A D and Muñoz-Elgueta, N and Nemmen, R S and Nilo Castellón, J L and Oliveira, A S and Ortíz, D and Pattaro, E and Pereira, C B and Quint, B and Riguccini, L and Rocha Pinto, H J and Rodrigues, I and Roig, F and Rossi, S and Saha, Kanak and Santos, R and Schnorr Müller, A and Sesto, Leandro A and Silva, R and Smith Castelli, Analia V and Teixeira, R and Telles, E and Thom de Souza, R C and Thöne, C and Trevisan, M and de Ugarte Postigo, A and Urrutia-Viscarra, F and Veiga, C H and Vika, M and Vitorelli, A Z and Werle, A and Werner, S V and Zaritsky, D},
    title = "{The Southern Photometric Local Universe Survey (S-PLUS): improved SEDs, morphologies, and redshifts with 12 optical filters}",
    journal = {Monthly Notices of the Royal Astronomical Society},
    volume = {489},
    number = {1},
    pages = {241-267},
    year = {2019},
    month = {08},
    issn = {0035-8711},
    doi = {10.1093/mnras/stz1985},
}

@ARTICLE{aladin-desktop,
       author = {{Bonnarel}, F. and {Fernique}, P. and {Bienaym{\'e}}, O. and {Egret}, D. and {Genova}, F. and {Louys}, M. and {Ochsenbein}, F. and {Wenger}, M. and {Bartlett}, J.~G.},
        title = "{The ALADIN interactive sky atlas. A reference tool for identification of astronomical sources}",
      journal = {Astronomy and Astrophysics, Supplement},
         year = 2000,
        month = apr,
       volume = {143},
        pages = {33-40},
          doi = {10.1051/aas:2000331},
       adsurl = {https://ui.adsabs.harvard.edu/abs/2000A&AS..143...33B}
}

@MISC{scs,
       author = {{Plante}, Raymond and {Williams}, Roy and {Hanisch}, Robert and {Szalay}, Alex},
        title = "{Simple Cone Search Version 1.03}",
 howpublished = {IVOA Recommendation 22 February 2008},
         year = 2008,
        month = feb,
        pages = {222},
          doi = {10.5479/ADS/bib/2008ivoa.specQ0222P},
archivePrefix = {arXiv},
       eprint = {1110.0498},
 primaryClass = {astro-ph.IM},
       adsurl = {https://ui.adsabs.harvard.edu/abs/2008ivoa.specQ0222P}
}

@inproceedings{skyserver,
author = {Szalay, Alexander S. and Gray, Jim and Thakar, Ani R. and Kunszt, Peter Z. and Malik, Tanu and Raddick, Jordan and Stoughton, Christopher and vandenBerg, Jan},
title = {The SDSS skyserver: public access to the sloan digital sky server data},
year = {2002},
isbn = {1581134975},
publisher = {Association for Computing Machinery},
address = {New York, NY, USA},
url = {https://doi.org/10.1145/564691.564758},
doi = {10.1145/564691.564758},
booktitle = {Proceedings of the 2002 ACM SIGMOD International Conference on Management of Data},
pages = {570–581},
numpages = {12},
location = {Madison, Wisconsin},
series = {SIGMOD '02}
}

@article{sciserver,
title = {SciServer: A science platform for astronomy and beyond},
journal = {Astronomy and Computing},
volume = {33},
pages = {100412},
year = {2020},
issn = {2213-1337},
doi = {https://doi.org/10.1016/j.ascom.2020.100412},
url = {https://doi.org/10.1016/j.ascom.2020.100412},
author = {M. Taghizadeh-Popp and J.W. Kim and G. Lemson and D. Medvedev and M.J. Raddick and A.S. Szalay and A.R. Thakar and J. Booker and C. Chhetri and L. Dobos and M. Rippin},
}

@MISC{ri,
       author = {{Dower}, Theresa and {Demleitner}, Markus and {Benson}, Kevin and {Plante}, Ray and {Auden}, Elizabeth and {Graham}, Matthew and {Greene}, Gretchen and {Hill}, Martin and {Linde}, Tony and {Morris}, Dave and {O`Mullane}, Wil and {Rixon}, Guy and {St{\'e}b{\'e}}, Aur{\'e}lien and {Andrews}, Kona},
        title = "{Registry Interfaces Version 1.1}",
 howpublished = {IVOA Recommendation 23 July 2018},
         year = 2018,
        month = jul,
        pages = {723},
          doi = {10.5479/ADS/bib/2018ivoa.spec.0723D},
       adsurl = {https://ui.adsabs.harvard.edu/abs/2018ivoa.spec.0723D}
}

@article{samp,
   author = {M. B. Taylor and T. Boch and J. Taylor},
   doi = {10.1016/J.ASCOM.2014.12.007},
   issn = {22131337},
   issue = {PB},
   journal = {Astronomy and Computing},
   month = {6},
   pages = {81-90},
   publisher = {Elsevier},
   title = {SAMP, the Simple Application Messaging Protocol: Letting applications talk to each other},
   volume = {11},
   url = {https://doi.org/10.1016/J.ASCOM.2014.12.007},
   year = {2015},
}

@MISC{samp-ivoa,
       author = {{Boch}, T. and {Fitzpatrick}, M. and {Taylor}, M. and {Allan}, A. and {Fay}, J. and {Paioro}, L. and {Taylor}, J. and {Tody}, D.},
        title = "{Simple Application Messaging Protocol Version 1.3}",
 howpublished = {IVOA Recommendation 11 April 2012},
         year = 2012,
        month = apr,
        pages = {411},
          doi = {10.5479/ADS/bib/2012ivoa.spec.0411B},
archivePrefix = {arXiv},
       eprint = {1110.0528},
 primaryClass = {astro-ph.IM},
       adsurl = {https://ui.adsabs.harvard.edu/abs/2012ivoa.spec.0411B}
}

@misc{matplotlib, 
title={Matplotlib: A 2D Graphics Environment}, 
volume={9}, 
url={http://doi.org/10.1109/MCSE.2007.55}, 
DOI={10.1109/mcse.2007.55}, 
number={3}, 
journal={Computing in Science &amp; Engineering}, 
publisher={Institute of Electrical and Electronics Engineers (IEEE)}, 
author={Hunter, John D.}, 
year={2007}, 
pages={90–95} 
}

@article{mar,
  title = {MAR: A Multiband Astronomical Reduction package},
  volume = {51},
  ISSN = {2213-1337},
  url = {http://doi.org/10.1016/j.ascom.2024.100899},
  DOI = {10.1016/j.ascom.2024.100899},
  journal = {Astronomy and Computing},
  publisher = {Elsevier BV},
  author = {Schwarz,  G.B. Oliveira and Herpich,  F. and Almeida-Fernandes,  F. and Nakazono,  L. and Cardoso,  N.M. and Machado-Pereira,  E. and Schoenell,  W. and Perottoni,  H.D. and Menéndez-Delmestre,  K. and Sodré,  L. and Kanaan,  A. and Ribeiro,  T.},
  year = {2025},
  month = apr,
  pages = {100899}
}

@article{astromorphlib,
  title = {Diagnostic diagrams for ram pressure stripped candidates},
  volume = {528},
  ISSN = {1365-2966},
  url = {http://doi.org/10.1093/mnras/stad3881},
  DOI = {10.1093/mnras/stad3881},
  number = {2},
  journal = {Monthly Notices of the Royal Astronomical Society},
  publisher = {Oxford University Press (OUP)},
  author = {Krabbe,  A C and Hernandez-Jimenez,  J A and Mendes de Oliveira,  C and Jaffe,  Y L and Oliveira,  C B and Cardoso,  N M and Smith Castelli,  A V and Dors,  O L and Cortesi,  A and Crossett,  J P},
  year = {2023},
  month = dec,
  pages = {1125–1141}
}

@article{analia,
  title = {The S-PLUS Fornax Project (S+FP): A first 12-band glimpse of the Fornax galaxy cluster},
  volume = {530},
  ISSN = {1365-2966},
  url = {http://doi.org/10.1093/mnras/stae840},
  DOI = {10.1093/mnras/stae840},
  number = {4},
  journal = {Monthly Notices of the Royal Astronomical Society},
  publisher = {Oxford University Press (OUP)},
  author = {Smith Castelli,  A V and Cortesi,  A and Haack,  R F and Lopes,  A R and Thainá-Batista,  J and Cid Fernandes,  R and Lomelí-Núñez,  L and Ribeiro,  U and de Bom,  C R and Cernic,  V and Sodré Jr,  L and Zenocratti,  L and De Rossi,  M E and Calderón,  J P and Herpich,  F and Telles,  E and Saha,  K and Lopes,  P A A and Lopes-Silva,  V H and Gon\c{c}alves,  T S and Bambrila,  D and Cardoso,  N M and Buzzo,  M L and Astudillo Sotomayor,  P and Demarco,  R and Leigh,  N and Sarzi,  M and Menéndez-Delmestre,  K and Faifer,  F R and Jiménez-Teja,  Y and Grossi,  M and Hernández-Jiménez,  J A and Krabbe,  A C and Gutiérrez Soto,  L A and Brandão,  D and Espinosa,  L and Olave-Rojas,  D E and Oliveira Schwarz,  G B and Almeida-Fernandes,  F and Schoenell,  W and Ribeiro,  T and Kanaan,  A and Mendes de Oliveira,  C},
  year = {2024},
  month = apr,
  pages = {3787–3811}
}

@article{clecio-1,
  title = {Deep Learning assessment of galaxy morphology in S-PLUS Data Release 1},
  volume = {507},
  ISSN = {1365-2966},
  url = {http://doi.org/10.1093/mnras/stab1981},
  DOI = {10.1093/mnras/stab1981},
  number = {2},
  journal = {Monthly Notices of the Royal Astronomical Society},
  publisher = {Oxford University Press (OUP)},
  author = {Bom,  C R and Cortesi,  A and Lucatelli,  G and Dias,  L O and Schubert,  P and Oliveira Schwarz,  G B and Cardoso,  N M and Lima,  E V R and Mendes de Oliveira,  C and Sodre,  L and Smith Castelli,  A V and Ferrari,  F and Damke,  G and Overzier,  R and Kanaan,  A and Ribeiro,  T and Schoenell,  W},
  year = {2021},
  month = jul,
  pages = {1937–1955}
}

@article{natanael,
  title = {Classifica\c{c}ão Morfológica de Galáxias no S-PLUS por Combina\c{c}ão de Redes Convolucionais},
  volume = {11},
  ISSN = {2236-7640},
  url = {http://doi.org/10.7437/nt2236-7640/2021.02.002},
  DOI = {10.7437/nt2236-7640/2021.02.002},
  number = {2},
  journal = {Notas Técnicas},
  publisher = {Brazilian Center for Physical Research},
  author = {Cardoso,  N. M. and Schwarz,  G. B. and Dias,  L. O. and Bom,  C. R. and Jr.,  L. Sodré and de Oliveira,  C. Mendes},
  year = {2021},
  month = oct,
  pages = {1–18}
}

@inproceedings{stilts,
  title = {{{STILTS}} - {{A Package}} for {{Command-Line Processing}} of {{Tabular Data}}},
  booktitle = {Astronomical {{Data Analysis Software}} and {{Systems XV}}},
  author = {Taylor, M. B.},
  year = {2006},
  month = jul,
  series = {Astronomical {{Society}} of the {{Pacific Conference Series}}},
  volume = {351},
  pages = {666},
  address = {San Lorenzo de El Escorial, Spain},
  url = {https://ui.adsabs.harvard.edu/abs/2006ASPC..351..666T},
  urldate = {2025-01-18}
}

@inproceedings{specview-2,
  title = {Specview: A {{Java Tool}} for {{Spectral Visualization}} and {{Model Fitting}}},
  shorttitle = {Specview},
  booktitle = {Astronomical {{Data Analysis Software}} and {{Systems XI}}},
  author = {Busko, I.},
  year = {2002},
  month = jan,
  series = {Astronomical {{Society}} of the {{Pacific Conference Series}}},
  volume = {281},
  pages = {120},
  url = {https://ui.adsabs.harvard.edu/abs/2002ASPC..281..120B},
  urldate = {2025-01-18},
  isbn = {1-58381-124-9}
}

@inproceedings{specview,
  title = {{{SPECVIEW}}: {{An Interactive Java Tool}} for {{Visualization}} and {{Analysis}} of {{Spectral Data}}},
  shorttitle = {{{SPECVIEW}}},
  booktitle = {Astronomical {{Data Analysis Software}} and {{Systems IX}}},
  author = {Busko, I.},
  year = {2000},
  series = {Astronomical {{Society}} of the {{Pacific Conference Series}}},
  volume = {216},
  pages = {79},
  url = {https://ui.adsabs.harvard.edu/abs/2000ASPC..216...79B},
  urldate = {2025-01-18},
  isbn = {1-58381-047-1}
}

@article{splat,
  title = {{{SPLAT}}: {{Spectral Analysis Tool}}},
  shorttitle = {{{SPLAT}}},
  author = {Draper, Peter W.},
  year = {2014},
  month = feb,
  journal = {Astrophysics Source Code Library},
  pages = {ascl:1402.007},
  url = {https://ui.adsabs.harvard.edu/abs/2014ascl.soft02007D},
  urldate = {2025-01-18}
}

@article{splatvo,
  title = {Spectroscopic Analysis in the Virtual Observatory Environment with {{SPLAT-VO}}},
  author = {{\v S}koda, P. and Draper, P. W. and Neves, M. C. and Andre{\v s}i{\v c}, D. and Jenness, T.},
  year = {2014},
  month = nov,
  journal = {Astronomy and Computing},
  series = {Special {{Issue}} on {{The Virtual Observatory}}: {{I}}},
  volume = {7--8},
  pages = {108--120},
  issn = {2213-1337},
  doi = {10.1016/j.ascom.2014.06.001},
  url = {https://www.sciencedirect.com/science/article/pii/S2213133714000250},
  urldate = {2025-01-18}
}

@inproceedings{splatvo-timeseries,
  title = {Time Series, Collaboration and Large Data Sets Enhancements of {{SPLAT-VO}}},
  booktitle = {2016 International Conference on Systems Informatics, Modelling and Simulation ({{SIMS}})},
  author = {{\u S}aloun, Petr and Andre{\u s}i{\u c}, David and {\u S}koda, Petr and Zelinka, Ivan},
  year = {2016},
  pages = {111--116},
  address = {Riga, Latvia},
  doi = {10.1109/SIMS.2016.20},
  isbn = {978-1-5090-2693-7}
}

@article{topcat-stil,
  title = {{{TOPCAT}} \& {{STIL}}: {{Starlink Table}}/{{VOTable Processing Software}}},
  author = {Taylor, M. B.},
  year = {2005},
  journal = {ASPC},
  volume = {347},
  pages = {29},
  issn = {1050-3390},
  url = {https://ui.adsabs.harvard.edu/abs/2005ASPC..347...29T/abstract},
  urldate = {2024-10-12}
}

@article{astropy2013,
  title = {Astropy: {{A}} Community {{Python}} Package for Astronomy},
  shorttitle = {Astropy},
  author = {{The Astropy Collaboration} and Robitaille, Thomas P. and Tollerud, Erik J. and Greenfield, Perry and Droettboom, Michael and Bray, Erik and Aldcroft, Tom and Davis, Matt and Ginsburg, Adam and {Price-Whelan}, Adrian M. and Kerzendorf, Wolfgang E. and Conley, Alexander and Crighton, Neil and Barbary, Kyle and Muna, Demitri and Ferguson, Henry and Grollier, Fr{\'e}d{\'e}ric and Parikh, Madhura M. and Nair, Prasanth H. and G{\"u}nther, Hans M. and Deil, Christoph and Woillez, Julien and Conseil, Simon and Kramer, Roban and Turner, James E. H. and Singer, Leo and Fox, Ryan and Weaver, Benjamin A. and Zabalza, Victor and Edwards, Zachary I. and Azalee Bostroem, K. and Burke, D. J. and Casey, Andrew R. and Crawford, Steven M. and Dencheva, Nadia and Ely, Justin and Jenness, Tim and Labrie, Kathleen and Lim, Pey Lian and Pierfederici, Francesco and Pontzen, Andrew and Ptak, Andy and Refsdal, Brian and Servillat, Mathieu and Streicher, Ole},
  year = {2013},
  month = oct,
  journal = {Astronomy \& Astrophysics},
  volume = {558},
  pages = {A33},
  issn = {0004-6361, 1432-0746},
  doi = {10.1051/0004-6361/201322068},
  url = {http://www.aanda.org/10.1051/0004-6361/201322068},
  urldate = {2025-01-28}
}

@article{astropy2018,
  title = {The {{Astropy Project}}: {{Building}} an {{Open-science Project}} and {{Status}} of the v2.0 {{Core Package}}{\textsuperscript{*}}},
  shorttitle = {The {{Astropy Project}}},
  author = {{The Astropy Collaboration} and {Price-Whelan}, A. M. and Sip{\H o}cz, B. M. and G{\"u}nther, H. M. and Lim, P. L. and Crawford, S. M. and Conseil, S. and Shupe, D. L. and Craig, M. W. and Dencheva, N. and Ginsburg, A. and VanderPlas, J. T. and Bradley, L. D. and {P{\'e}rez-Su{\'a}rez}, D. and {De Val-Borro}, M. and {(Primary Paper Contributors)} and Aldcroft, T. L. and Cruz, K. L. and Robitaille, T. P. and Tollerud, E. J. and {(Astropy Coordination Committee)} and Ardelean, C. and Babej, T. and Bach, Y. P. and Bachetti, M. and Bakanov, A. V. and Bamford, S. P. and Barentsen, G. and Barmby, P. and Baumbach, A. and Berry, K. L. and Biscani, F. and Boquien, M. and Bostroem, K. A. and Bouma, L. G. and Brammer, G. B. and Bray, E. M. and Breytenbach, H. and Buddelmeijer, H. and Burke, D. J. and Calderone, G. and Rodr{\'i}guez, J. L. Cano and Cara, M. and Cardoso, J. V. M. and Cheedella, S. and Copin, Y. and Corrales, L. and Crichton, D. and D'Avella, D. and Deil, C. and Depagne, {\'E}. and Dietrich, J. P. and Donath, A. and Droettboom, M. and Earl, N. and Erben, T. and Fabbro, S. and Ferreira, L. A. and Finethy, T. and Fox, R. T. and Garrison, L. H. and Gibbons, S. L. J. and Goldstein, D. A. and Gommers, R. and Greco, J. P. and Greenfield, P. and Groener, A. M. and Grollier, F. and Hagen, A. and Hirst, P. and Homeier, D. and Horton, A. J. and Hosseinzadeh, G. and Hu, L. and Hunkeler, J. S. and Ivezi{\'c}, {\v Z}. and Jain, A. and Jenness, T. and Kanarek, G. and Kendrew, S. and Kern, N. S. and Kerzendorf, W. E. and Khvalko, A. and King, J. and Kirkby, D. and Kulkarni, A. M. and Kumar, A. and Lee, A. and Lenz, D. and Littlefair, S. P. and Ma, Z. and Macleod, D. M. and Mastropietro, M. and McCully, C. and Montagnac, S. and Morris, B. M. and Mueller, M. and Mumford, S. J. and Muna, D. and Murphy, N. A. and Nelson, S. and Nguyen, G. H. and Ninan, J. P. and N{\"o}the, M. and Ogaz, S. and Oh, S. and Parejko, J. K. and Parley, N. and Pascual, S. and Patil, R. and Patil, A. A. and Plunkett, A. L. and Prochaska, J. X. and Rastogi, T. and Janga, V. Reddy and Sabater, J. and Sakurikar, P. and Seifert, M. and Sherbert, L. E. and {Sherwood-Taylor}, H. and Shih, A. Y. and Sick, J. and Silbiger, M. T. and Singanamalla, S. and Singer, L. P. and Sladen, P. H. and Sooley, K. A. and Sornarajah, S. and Streicher, O. and Teuben, P. and Thomas, S. W. and Tremblay, G. R. and Turner, J. E. H. and Terr{\'o}n, V. and Kerkwijk, M. H. Van and De La Vega, A. and Watkins, L. L. and Weaver, B. A. and Whitmore, J. B. and Woillez, J. and Zabalza, V. and {(Astropy Contributors)}},
  year = {2018},
  month = sep,
  journal = {The Astronomical Journal},
  volume = {156},
  number = {3},
  pages = {123},
  issn = {0004-6256, 1538-3881},
  doi = {10.3847/1538-3881/aabc4f},
  url = {https://iopscience.iop.org/article/10.3847/1538-3881/aabc4f},
  urldate = {2025-01-28}
}

@article{astropy2022,
  title = {The {{Astropy Project}}: {{Sustaining}} and {{Growing}} a {{Community-oriented Open-source Project}} and the {{Latest Major Release}} (v5.0) of the {{Core Package}}*},
  shorttitle = {The {{Astropy Project}}},
  author = {{The Astropy Collaboration} and {Price-Whelan}, Adrian M. and Lim, Pey Lian and Earl, Nicholas and Starkman, Nathaniel and Bradley, Larry and Shupe, David L. and Patil, Aarya A. and Corrales, Lia and Brasseur, C. E. and N{\"o}the, Maximilian and Donath, Axel and Tollerud, Erik and Morris, Brett M. and Ginsburg, Adam and Vaher, Eero and Weaver, Benjamin A. and Tocknell, James and Jamieson, William and Van Kerkwijk, Marten H. and Robitaille, Thomas P. and Merry, Bruce and Bachetti, Matteo and G{\"u}nther, H. Moritz and {Paper Authors} and Aldcroft, Thomas L. and {Alvarado-Montes}, Jaime A. and Archibald, Anne M. and B{\'o}di, Attila and Bapat, Shreyas and Barentsen, Geert and Baz{\'a}n, Juanjo and Biswas, Manish and Boquien, M{\'e}d{\'e}ric and Burke, D. J. and Cara, Daria and Cara, Mihai and Conroy, Kyle E and Conseil, Simon and Craig, Matthew W. and Cross, Robert M. and Cruz, Kelle L. and D'Eugenio, Francesco and Dencheva, Nadia and Devillepoix, Hadrien A. R. and Dietrich, J{\"o}rg P. and Eigenbrot, Arthur Davis and Erben, Thomas and Ferreira, Leonardo and {Foreman-Mackey}, Daniel and Fox, Ryan and Freij, Nabil and Garg, Suyog and Geda, Robel and Glattly, Lauren and Gondhalekar, Yash and Gordon, Karl D. and Grant, David and Greenfield, Perry and Groener, Austen M. and Guest, Steve and Gurovich, Sebastian and Handberg, Rasmus and Hart, Akeem and {Hatfield-Dodds}, Zac and Homeier, Derek and Hosseinzadeh, Griffin and Jenness, Tim and Jones, Craig K. and Joseph, Prajwel and Kalmbach, J. Bryce and Karamehmetoglu, Emir and Ka{\l}uszy{\'n}ski, Miko{\l}aj and Kelley, Michael S. P. and Kern, Nicholas and Kerzendorf, Wolfgang E. and Koch, Eric W. and Kulumani, Shankar and Lee, Antony and Ly, Chun and Ma, Zhiyuan and MacBride, Conor and Maljaars, Jakob M. and Muna, Demitri and Murphy, N. A. and Norman, Henrik and O'Steen, Richard and Oman, Kyle A. and Pacifici, Camilla and Pascual, Sergio and {Pascual-Granado}, J. and Patil, Rohit R. and Perren, Gabriel I and Pickering, Timothy E. and Rastogi, Tanuj and Roulston, Benjamin R. and Ryan, Daniel F and Rykoff, Eli S. and Sabater, Jose and Sakurikar, Parikshit and Salgado, Jes{\'u}s and Sanghi, Aniket and Saunders, Nicholas and Savchenko, Volodymyr and Schwardt, Ludwig and {Seifert-Eckert}, Michael and Shih, Albert Y. and Jain, Anany Shrey and Shukla, Gyanendra and Sick, Jonathan and Simpson, Chris and Singanamalla, Sudheesh and Singer, Leo P. and Singhal, Jaladh and Sinha, Manodeep and Sip{\H o}cz, Brigitta M. and Spitler, Lee R. and Stansby, David and Streicher, Ole and {\v S}umak, Jani and Swinbank, John D. and Taranu, Dan S. and Tewary, Nikita and Tremblay, Grant R. and {Val-Borro}, Miguel De and Van Kooten, Samuel J. and Vasovi{\'c}, Zlatan and Verma, Shresth and De Miranda Cardoso, Jos{\'e} Vin{\'i}cius and Williams, Peter K. G. and Wilson, Tom J. and Winkel, Benjamin and {Wood-Vasey}, W. M. and Xue, Rui and Yoachim, Peter and Zhang, Chen and Zonca, Andrea and {Astropy Project Contributors}},
  year = {2022},
  month = aug,
  journal = {The Astrophysical Journal},
  volume = {935},
  number = {2},
  pages = {167},
  issn = {0004-637X, 1538-4357},
  doi = {10.3847/1538-4357/ac7c74},
  url = {https://iopscience.iop.org/article/10.3847/1538-4357/ac7c74},
  urldate = {2025-01-28}
}

@article{gz,
  title = {Galaxy {{Zoo}}: {{Morphologies}} Derived from Visual Inspection of Galaxies from the {{Sloan Digital Sky Survey}}},
  author = {Lintott, Chris J. and Schawinski, Kevin and Slosar, An{\v z}e and Land, Kate and Bamford, Steven and Thomas, Daniel and Raddick, M. Jordan and Nichol, Robert C. and Szalay, Alex and Andreescu, Dan and Murray, Phil and Vandenberg, Jan},
  year = {2008},
  month = sep,
  journal = {Monthly Notices of the Royal Astronomical Society},
  volume = {389},
  number = {3},
  eprint = {0804.4483},
  pages = {1179--1189},
  publisher = {Oxford University Press},
  issn = {13652966},
  doi = {10.1111/J.1365-2966.2008.13689.X},
  url = {https://doi.org/10.1111/j.1365-2966.2008.13689.x},
  urldate = {2024-11-10},
  archiveprefix = {arXiv}
}

@article{gz1,
  title = {Galaxy {{Zoo}} 1: Data Release of Morphological Classifications for Nearly 900 000 Galaxies},
  author = {Lintott, Chris and Schawinski, Kevin and Bamford, Steven and Slosar, An{\v z}e and Land, Kate and Thomas, Daniel and Edmondson, Edd and Masters, Karen and Nichol, Robert C. and Raddick, M. Jordan and Szalay, Alex and Andreescu, Dan and Murray, Phil and Vandenberg, Jan},
  year = {2011},
  month = jan,
  journal = {Monthly Notices of the Royal Astronomical Society},
  volume = {410},
  number = {1},
  eprint = {1007.3265},
  pages = {166--178},
  publisher = {Oxford Academic},
  issn = {0035-8711},
  doi = {10.1111/J.1365-2966.2010.17432.X},
  url = {https://doi.org/10.1111/j.1365-2966.2010.17432.x},
  urldate = {2024-11-10},
  archiveprefix = {arXiv}
}

@article{abell-clusters,
  title = {A Catalog of Rich Clusters of Galaxies},
  author = {Abell, George O. and Corwin, Jr., Harold G. and Olowin, Ronald P.},
  year = {1989},
  month = may,
  journal = {The Astrophysical Journal Supplement Series},
  volume = {70},
  pages = {1},
  issn = {0067-0049, 1538-4365},
  doi = {10.1086/191333},
  url = {http://adsabs.harvard.edu/doi/10.1086/191333},
  urldate = {2025-01-28},
  langid = {english},
  language = {en}
}

@misc{erik-redshifts,
  doi = {10.5281/ZENODO.15127060},
  url = {https://zenodo.org/doi/10.5281/zenodo.15127060},
  author = {Lima, Erik V},
  title = {Southern Hemisphere Spectrocopic Redshift Compilation},
  publisher = {Zenodo},
  year = {2025},
  copyright = {MIT License}
}

@incollection{producer-consumer68,
  title = {Cooperating {{Sequential Processes}}},
  booktitle = {The {{Origin}} of {{Concurrent Programming}}},
  author = {Dijkstra, Edsger W.},
  editor = {Hansen, Per Brinch},
  year = {1968},
  pages = {65--138},
  publisher = {Springer New York},
  address = {New York, NY},
  doi = {10.1007/978-1-4757-3472-0_2},
  url = {http://link.springer.com/10.1007/978-1-4757-3472-0\_2},
  urldate = {2025-02-05},
  isbn = {978-1-4419-2986-0 978-1-4757-3472-0},
  langid = {english},
  language = {en}
}

@article{producer-consumer72,
  title = {Information Streams Sharing a Finite Buffer},
  author = {Dijkstra, Edsger W.},
  year = {1972},
  month = oct,
  journal = {Information Processing Letters},
  volume = {1},
  number = {5},
  pages = {179--180},
  issn = {00200190},
  doi = {10.1016/0020-0190(72)90034-8},
  url = {https://linkinghub.elsevier.com/retrieve/pii/0020019072900348},
  urldate = {2025-02-05},
  copyright = {https://www.elsevier.com/tdm/userlicense/1.0/},
  langid = {english},
  language = {en}
}

@article{panagoulia2014,
	title = {A volume-limited sample of {X}-ray galaxy groups and clusters – {I}. {Radial} entropy and cooling time profiles},
	volume = {438},
	issn = {0035-8711, 1365-2966},
	url = {http://academic.oup.com/mnras/article/438/3/2341/971222/A-volumelimited-sample-of-Xray-galaxy-groups-and},
	doi = {10.1093/mnras/stt2349},
	language = {en},
	number = {3},
	urldate = {2025-08-08},
	journal = {Monthly Notices of the Royal Astronomical Society},
	author = {Panagoulia, E. K. and Fabian, A. C. and Sanders, J. S.},
	month = mar,
	year = {2014},
	pages = {2341--2354},
}

@article{bohringer2002,
	title = {The {Mass} {Function} of an {X}‐{Ray} {Flux}–limited {Sample} of {Galaxy} {Clusters}},
	volume = {567},
	issn = {0004-637X, 1538-4357},
	url = {https://iopscience.iop.org/article/10.1086/338753},
	doi = {10.1086/338753},
	language = {en},
	number = {2},
	urldate = {2025-08-08},
	journal = {The Astrophysical Journal},
	author = {Reiprich, Thomas H. and Bohringer, Hans},
	month = mar,
	year = {2002},
	pages = {716--740},
}

@article{jorgensen1996,
	title = {The {Fundamental} {Plane} for cluster {E} and {S0} galaxies},
	volume = {280},
	issn = {0035-8711, 1365-2966},
	url = {https://academic.oup.com/mnras/article-lookup/doi/10.1093/mnras/280.1.167},
	doi = {10.1093/mnras/280.1.167},
	language = {en},
	number = {1},
	urldate = {2025-08-08},
	journal = {Monthly Notices of the Royal Astronomical Society},
	author = {Jorgensen, I. and Franx, M. and Kjaergaard, P.},
	month = may,
	year = {1996},
	pages = {167--185},
}

@article{struble1999,
	title = {A {Compilation} of {Redshifts} and {Velocity} {Dispersions} for {ACO} {Clusters}},
	volume = {125},
	issn = {0067-0049, 1538-4365},
	url = {https://iopscience.iop.org/article/10.1086/313274},
	doi = {10.1086/313274},
	language = {en},
	number = {1},
	urldate = {2025-08-08},
	journal = {The Astrophysical Journal Supplement Series},
	author = {Struble, Mitchell F. and Rood, Herbert J.},
	month = nov,
	year = {1999},
	pages = {35--71},
}

@article{efigi,
	title = {The {EFIGI} catalogue of 4458 nearby galaxies with detailed morphology},
	volume = {532},
	issn = {0004-6361, 1432-0746},
	url = {http://www.aanda.org/10.1051/0004-6361/201016423},
	doi = {10.1051/0004-6361/201016423},
	urldate = {2025-08-04},
	journal = {Astronomy \& Astrophysics},
	author = {Baillard, A. and Bertin, E. and De Lapparent, V. and Fouqué, P. and Arnouts, S. and Mellier, Y. and Pelló, R. and Leborgne, J.-F. and Prugniel, P. and Makarov, D. and Makarova, L. and McCracken, H. J. and Bijaoui, A. and Tasca, L.},
	month = aug,
	year = {2011},
	pages = {A74},
}

@incollection{tssc-natural,
	address = {Cham},
	title = {Citizen {Science} in the {Natural} {Sciences}},
	isbn = {978-3-030-58278-4},
	url = {https://doi.org/10.1007/978-3-030-58278-4_5},
	booktitle = {The {Science} of {Citizen} {Science}},
	publisher = {Springer International Publishing},
	author = {Frigerio Didone
and Richter, Anett
and Per Esra
and Pruse Baiba
and Vohland Katrin},
	editor = {Vohland Katrin
and Land-Zandstra, Anne
and Ceccaroni Luigi
and Lemmens Rob
and Perelló Josep
and Ponti Marisa
and Samson Roeland
and Wagenknecht Katherin},
	year = {2021},
	doi = {10.1007/978-3-030-58278-4_5},
	pages = {79--96},
}

@incollection{tssc-platforms,
	address = {Cham},
	title = {Citizen {Science} {Platforms}},
	isbn = {978-3-030-58278-4},
	url = {https://doi.org/10.1007/978-3-030-58278-4_22},
	booktitle = {The {Science} of {Citizen} {Science}},
	publisher = {Springer International Publishing},
	author = {Liu Hai-Ying
and Dörler, Daniel
and Heigl Florian
and Grossberndt Sonja},
	editor = {Vohland Katrin
and Land-Zandstra, Anne
and Ceccaroni Luigi
and Lemmens Rob
and Perelló Josep
and Ponti Marisa
and Samson Roeland
and Wagenknecht Katherin},
	year = {2021},
	doi = {10.1007/978-3-030-58278-4_22},
	pages = {439--459},
}

@article{lsst,
	title = {{LSST}: {From} {Science} {Drivers} to {Reference} {Design} and {Anticipated} {Data} {Products}},
	volume = {873},
	issn = {0004-637X},
	url = {https://doi.org/10.3847/1538-4357/AB042C},
	doi = {10.3847/1538-4357/AB042C},
	number = {2},
	urldate = {2024-11-11},
	journal = {The Astrophysical Journal},
	author = {Ivezi{\'c}, Z{\v e}ljko and others},
	month = mar,
	year = {2019},
	note = {arXiv: 0805.2366
Publisher: IOP Publishing},
	pages = {111},
}

@article{nakazono2021,
	title = {On the discovery of stars, quasars, and galaxies in the {Southern} {Hemisphere} with {S}-{PLUS} {DR2}},
	volume = {507},
	copyright = {https://academic.oup.com/journals/pages/open\_access/funder\_policies/chorus/standard\_publication\_model},
	issn = {0035-8711, 1365-2966},
	url = {https://academic.oup.com/mnras/article/507/4/5847/6319514},
	doi = {10.1093/mnras/stab1835},
	language = {en},
	number = {4},
	urldate = {2025-09-10},
	journal = {Monthly Notices of the Royal Astronomical Society},
	author = {Nakazono, L and Mendes de Oliveira, C and Hirata, N S T and Jeram, S and Queiroz, C and Eikenberry, Stephen S and Gonzalez, A H and Abramo, R and Overzier, R and Espadoto, M and Martinazzo, A and Sampedro, L and Herpich, F R and Almeida-Fernandes, F and Werle, A and Barbosa, C E and Sodré Jr., L and Lima, E V and Buzzo, M L and Cortesi, A and Menéndez-Delmestre, K and Akras, S and Alvarez-Candal, Alvaro and Lopes, A R and Telles, E and Schoenell, W and Kanaan, A and Ribeiro, T},
	month = sep,
	year = {2021},
	pages = {5847--5868},
}

@article{lima2022,
	title = {Photometric redshifts for the {S}-{PLUS} {Survey}: {Is} machine learning up to the task?},
	volume = {38},
	issn = {22131337},
	shorttitle = {Photometric redshifts for the {S}-{PLUS} {Survey}},
	url = {https://linkinghub.elsevier.com/retrieve/pii/S2213133721000640},
	doi = {10.1016/j.ascom.2021.100510},
	language = {en},
	urldate = {2025-09-10},
	journal = {Astronomy and Computing},
	author = {Lima, E.V.R. and Sodré, L. and Bom, C.R. and Teixeira, G.S.M. and Nakazono, L. and Buzzo, M.L. and Queiroz, C. and Herpich, F.R. and Castellon, J.L. Nilo and Dantas, M.L.L. and Dors, O.L. and Souza, R.C. Thom De and Akras, S. and Jiménez-Teja, Y. and Kanaan, A. and Ribeiro, T. and Schoennell, W.},
	month = jan,
	year = {2022},
	pages = {100510},
}

@article{lima-dias2023,
	title = {Bulge–disc decomposition of the {Hydra} cluster galaxies in 12 bands},
	volume = {527},
	copyright = {https://creativecommons.org/licenses/by/4.0/},
	issn = {0035-8711, 1365-2966},
	url = {https://academic.oup.com/mnras/article/527/3/5792/7438897},
	doi = {10.1093/mnras/stad3571},
	language = {en},
	number = {3},
	urldate = {2025-09-10},
	journal = {Monthly Notices of the Royal Astronomical Society},
	author = {Lima-Dias, Ciria and Monachesi, Antonela and Torres-Flores, Sergio and Cortesi, Arianna and Hernández-Lang, Daniel and P. Montaguth, Gissel and Jiménez-Teja, Yolanda and Panda, Swayamtrupta and Menéndez-Delmestre, Karín and Gonçalves, Thiago S and Méndez-Hernández, Hugo and Telles, Eduardo and Dimauro, Paola and Bom, Clécio R and Mendes de Oliveira, Claudia and Kanaan, Antonio and Ribeiro, Tiago and Schoenell, William},
	month = nov,
	year = {2023},
	pages = {5792--5807},
}

@article{gunn1972,
	title = {On the {Infall} of {Matter} {Into} {Clusters} of {Galaxies} and {Some} {Effects} on {Their} {Evolution}},
	volume = {176},
	issn = {0004-637X, 1538-4357},
	url = {http://adsabs.harvard.edu/doi/10.1086/151605},
	doi = {10.1086/151605},
	language = {en},
	urldate = {2025-09-10},
	journal = {The Astrophysical Journal},
	author = {Gunn, James E. and Gott, Iii, J. Richard},
	month = aug,
	year = {1972},
	pages = {1},
}

@article{jaffe2015,
	title = {{BUDHIES} {II}: a phase-space view of {H} i gas stripping and star formation quenching in cluster galaxies},
	volume = {448},
	issn = {1365-2966, 0035-8711},
	shorttitle = {{BUDHIES} {II}},
	url = {http://academic.oup.com/mnras/article/448/2/1715/1061483/BUDHIES-II-a-phasespace-view-of-Hi-gas-stripping},
	doi = {10.1093/mnras/stv100},
	language = {en},
	number = {2},
	urldate = {2025-09-10},
	journal = {Monthly Notices of the Royal Astronomical Society},
	author = {Jaffé, Yara L. and Smith, Rory and Candlish, Graeme N. and Poggianti, Bianca M. and Sheen, Yun-Kyeong and Verheijen, Marc A. W.},
	month = apr,
	year = {2015},
	pages = {1715--1728},
}

@article{moore1996,
	title = {Galaxy harassment and the evolution of clusters of galaxies},
	volume = {379},
	copyright = {http://www.springer.com/tdm},
	issn = {0028-0836, 1476-4687},
	url = {https://www.nature.com/articles/379613a0},
	doi = {10.1038/379613a0},
	language = {en},
	number = {6566},
	urldate = {2025-09-10},
	journal = {Nature},
	author = {Moore, Ben and Katz, Neal and Lake, George and Dressler, Alan and Oemler, Augustus},
	month = feb,
	year = {1996},
	pages = {613--616},
}

@article{gallagher1972,
	title = {A {Note} on {Mass} {Loss} during {Collisions} between {Galaxies} a nd the {Formation} of {Giant} {Systems}},
	volume = {77},
	issn = {00046256},
	url = {http://adsabs.harvard.edu/cgi-bin/bib_query?1972AJ.....77..288G},
	doi = {10.1086/111280},
	urldate = {2025-09-10},
	journal = {The Astronomical Journal},
	author = {Gallagher, Iii, John S. and Ostriker, Jeremiah P.},
	month = may,
	year = {1972},
	pages = {288},
}

@article{lima-dias2020,
	title = {An environmental dependence of the physical and structural properties in the {Hydra} cluster galaxies},
	volume = {500},
	copyright = {https://academic.oup.com/journals/pages/open\_access/funder\_policies/chorus/standard\_publication\_model},
	issn = {0035-8711, 1365-2966},
	url = {https://academic.oup.com/mnras/article/500/1/1323/5941516},
	doi = {10.1093/mnras/staa3326},
	language = {en},
	number = {1},
	urldate = {2025-09-11},
	journal = {Monthly Notices of the Royal Astronomical Society},
	author = {Lima-Dias, Ciria and Monachesi, Antonela and Torres-Flores, Sergio and Cortesi, Arianna and Hernández-Lang, Daniel and Barbosa, Carlos Eduardo and Mendes de Oliveira, Claudia and Olave-Rojas, Daniela and Pallero, Diego and Sampedro, Laura and Molino, Alberto and Herpich, Fabio R and Jaffé, Yara L and Amorín, Ricardo and Chies-Santos, Ana L and Dimauro, Paola and Telles, Eduardo and Lopes, Paulo A A and Alvarez-Candal, Alvaro and Ferrari, Fabricio and Kanaan, Antonio and Ribeiro, Tiago and Schoenell, William},
	month = nov,
	year = {2020},
	pages = {1323--1339},
}

@article{paudel2018,
	title = {A {Catalog} of {Merging} {Dwarf} {Galaxies} in the {Local} {Universe}},
	volume = {237},
	issn = {0067-0049},
	url = {https://iopscience.iop.org/article/10.3847/1538-4365/aad555},
	doi = {10.3847/1538-4365/AAD555},
	number = {2},
	urldate = {2022-01-04},
	journal = {The Astrophysical Journal Supplement Series},
	author = {Paudel, Sanjaya and Smith, Rory and Yoon, Suk Jin and Calderón-Castillo, Paula and Duc, Pierre-Alain},
	month = aug,
	year = {2018},
	note = {arXiv: 1807.07195
Publisher: IOP Publishing},
	pages = {36},
}

@article{galdeano2025,
	title = {Enlightening the {Universe} behind the {Milky} {Way} bulge: {II}. {Morphological} classification and galaxy properties},
	volume = {698},
	copyright = {https://creativecommons.org/licenses/by/4.0},
	issn = {0004-6361, 1432-0746},
	shorttitle = {Enlightening the {Universe} behind the {Milky} {Way} bulge},
	url = {https://www.aanda.org/10.1051/0004-6361/202554268},
	doi = {10.1051/0004-6361/202554268},
	urldate = {2025-07-18},
	journal = {Astronomy \& Astrophysics},
	author = {Galdeano, Daniela and Coldwell, Georgina and Alonso, Sol and Duplancic, Fernanda and Lucas, Philip W. and Fernandez, Julia and Perez, Noelia and Pereyra, Luis and Mesa, Valeria and Minniti, Dante and Smith, Leigh C. and Zarate, Francisco M.},
	month = jun,
	year = {2025},
	note = {Publisher: EDP Sciences},
	pages = {A214},
}

@article{tan2025,
	title = {A {Catalog} of {H} \textit{α} {Emission}-line {Stars} and 785 {Newly} {Identified} {Young} {Stellar} {Object} {Candidates} from {LAMOST} {Based} on a {Deep} {Learning} {Method}},
	volume = {280},
	issn = {0067-0049, 1538-4365},
	url = {https://iopscience.iop.org/article/10.3847/1538-4365/adf4e6},
	doi = {10.3847/1538-4365/adf4e6},
	number = {1},
	urldate = {2025-10-21},
	journal = {The Astrophysical Journal Supplement Series},
	author = {Tan, Lei and Mei, Ying and Qian, Jiale  and Wang , Xiaolong  and Cui, Yue and Huang , Aixin  and Wang , Feng  and Deng , Hui  and Liu , Chao  and Chi, Huanbin},
	month = sep,
	year = {2025},
	pages = {24},
}

@article{lambas2012,
	title = {Astrophysics {Galaxy} interactions {I}. {Major} and minor mergers},
	volume = {539},
	doi = {10.1051/0004-6361/201117900},
	urldate = {2021-11-21},
	journal = {A\&A},
	author = {Lambas, D G and Alonso, S and Mesa, V and O'mill, A L},
	year = {2012},
	pages = {45},
}

@article{darg2010b,
	title = {Galaxy {Zoo}: the properties of merging galaxies in the nearby {Universe} - local environments, colours, masses, star formation rates and {AGN} activity},
	volume = {401},
	issn = {00358711, 13652966},
	shorttitle = {Galaxy {Zoo}},
	url = {https://academic.oup.com/mnras/article-lookup/doi/10.1111/j.1365-2966.2009.15786.x},
	doi = {10.1111/j.1365-2966.2009.15786.x},
	language = {en},
	number = {3},
	urldate = {2025-08-04},
	journal = {Monthly Notices of the Royal Astronomical Society},
	author = {Darg, D. W. and Kaviraj, S. and Lintott, C. J. and Schawinski, K. and Sarzi, M. and Bamford, S. and Silk, J. and Andreescu, D. and Murray, P. and Nichol, R. C. and Raddick, M. J. and Slosar, A. and Szalay, A. S. and Thomas, D. and Vandenberg, J.},
	month = jan,
	year = {2010},
	pages = {1552--1563},
}

@article{des,
	title = {The {Dark} {Energy} {Survey}: more than dark energy - an overview},
	volume = {460},
	issn = {0035-8711},
	shorttitle = {The {Dark} {Energy} {Survey}},
	url = {https://ui.adsabs.harvard.edu/abs/2016MNRAS.460.1270D},
	doi = {10.1093/mnras/stw641},
	urldate = {2025-11-05},
	journal = {Monthly Notices of the Royal Astronomical Society},
	author = {{Dark Energy Survey Collaboration} and Abbott, T. and Abdalla, F. B. and Aleksić, J. and Allam, S. and Amara, A. and Bacon, D. and Balbinot, E. and Banerji, M. and Bechtol, K. and Benoit-Lévy, A. and Bernstein, G. M. and Bertin, E. and Blazek, J. and Bonnett, C. and Bridle, S. and Brooks, D. and Brunner, R. J. and Buckley-Geer, E. and Burke, D. L. and Caminha, G. B. and Capozzi, D. and Carlsen, J. and Carnero-Rosell, A. and Carollo, M. and Carrasco-Kind, M. and Carretero, J. and Castander, F. J. and Clerkin, L. and Collett, T. and Conselice, C. and Crocce, M. and Cunha, C. E. and D'Andrea, C. B. and da Costa, L. N. and Davis, T. M. and Desai, S. and Diehl, H. T. and Dietrich, J. P. and Dodelson, S. and Doel, P. and Drlica-Wagner, A. and Estrada, J. and Etherington, J. and Evrard, A. E. and Fabbri, J. and Finley, D. A. and Flaugher, B. and Foley, R. J. and Fosalba, P. and Frieman, J. and García-Bellido, J. and Gaztanaga, E. and Gerdes, D. W. and Giannantonio, T. and Goldstein, D. A. and Gruen, D. and Gruendl, R. A. and Guarnieri, P. and Gutierrez, G. and Hartley, W. and Honscheid, K. and Jain, B. and James, D. J. and Jeltema, T. and Jouvel, S. and Kessler, R. and King, A. and Kirk, D. and Kron, R. and Kuehn, K. and Kuropatkin, N. and Lahav, O. and Li, T. S. and Lima, M. and Lin, H. and Maia, M. A. G. and Makler, M. and Manera, M. and Maraston, C. and Marshall, J. L. and Martini, P. and McMahon, R. G. and Melchior, P. and Merson, A. and Miller, C. J. and Miquel, R. and Mohr, J. J. and Morice-Atkinson, X. and Naidoo, K. and Neilsen, E. and Nichol, R. C. and Nord, B. and Ogando, R. and Ostrovski, F. and Palmese, A. and Papadopoulos, A. and Peiris, H. V. and Peoples, J. and Percival, W. J. and Plazas, A. A. and Reed, S. L. and Refregier, A. and Romer, A. K. and Roodman, A. and Ross, A. and Rozo, E. and Rykoff, E. S. and Sadeh, I. and Sako, M. and Sánchez, C. and Sanchez, E. and Santiago, B. and Scarpine, V. and Schubnell, M. and Sevilla-Noarbe, I. and Sheldon, E. and Smith, M. and Smith, R. C. and Soares-Santos, M. and Sobreira, F. and Soumagnac, M. and Suchyta, E. and Sullivan, M. and Swanson, M. and Tarle, G. and Thaler, J. and Thomas, D. and Thomas, R. C. and Tucker, D. and Vieira, J. D. and Vikram, V. and Walker, A. R. and Wechsler, R. H. and Weller, J. and Wester, W. and Whiteway, L. and Wilcox, H. and Yanny, B. and Zhang, Y. and Zuntz, J.},
	month = aug,
	year = {2016},
	note = {Publisher: OUP
ADS Bibcode: 2016MNRAS.460.1270D},
	pages = {1270--1299},
}

@article{denis,
	title = {The deep near-infrared southern sky survey ({DENIS}).},
	volume = {87},
	issn = {0722-6691},
	url = {https://ui.adsabs.harvard.edu/abs/1997Msngr..87...27E},
	urldate = {2025-11-05},
	journal = {The Messenger},
	author = {Epchtein, N. and de Batz, B. and Capoani, L. and Chevallier, L. and Copet, E. and Fouqué, P. and Lacombe, P. and Le Bertre, T. and Pau, S. and Rouan, D. and Ruphy, S. and Simon, G. and Tiphène, D. and Burton, W. B. and Bertin, E. and Deul, E. and Habing, H. and Borsenberger, J. and Dennefeld, M. and Guglielmo, F. and Loup, C. and Mamon, G. and Ng, Y. and Omont, A. and Provost, L. and Renault, J.-C. and Tanguy, F. and Kimeswenger, S. and Kienel, C. and Garzon, F. and Persi, P. and Ferrari-Toniolo, M. and Robin, A. and Paturel, G. and Vauglin, I. and Forveille, T. and Delfosse, X. and Hron, J. and Schultheis, M. and Appenzeller, I. and Wagner, S. and Balazs, L. and Holl, A. and Lépine, J. and Boscolo, P. and Picazzio, E. and Duc, P.-A. and Mennessier, M.-O.},
	month = mar,
	year = {1997},
	note = {ADS Bibcode: 1997Msngr..87...27E},
	pages = {27--34},
}

@article{twomass,
	title = {The {Two} {Micron} {All} {Sky} {Survey} ({2MASS})},
	volume = {131},
	issn = {0004-6256, 1538-3881},
	url = {https://iopscience.iop.org/article/10.1086/498708},
	doi = {10.1086/498708},
	language = {en},
	number = {2},
	urldate = {2025-11-05},
	journal = {The Astronomical Journal},
	author = {Skrutskie, M. F. and Cutri, R. M. and Stiening, R. and Weinberg, M. D. and Schneider, S. and Carpenter, J. M. and Beichman, C. and Capps, R. and Chester, T. and Elias, J. and Huchra, J. and Liebert, J. and Lonsdale, C. and Monet, D. G. and Price, S. and Seitzer, P. and Jarrett, T. and Kirkpatrick, J. D. and Gizis, J. E. and Howard, E. and Evans, T. and Fowler, J. and Fullmer, L. and Hurt, R. and Light, R. and Kopan, E. L. and Marsh, K. A. and McCallon, H. L. and Tam, R. and Van Dyk, S. and Wheelock, S.},
	month = feb,
	year = {2006},
	pages = {1163--1183},
}

@article{wise,
	title = {{THE} {WIDE}-{FIELD} {INFRARED} {SURVEY} {EXPLORER} ({WISE}): {MISSION} {DESCRIPTION} {AND} {INITIAL} {ON}-{ORBIT} {PERFORMANCE}},
	volume = {140},
	issn = {0004-6256, 1538-3881},
	shorttitle = {{THE} {WIDE}-{FIELD} {INFRARED} {SURVEY} {EXPLORER} ({WISE})},
	url = {https://iopscience.iop.org/article/10.1088/0004-6256/140/6/1868},
	doi = {10.1088/0004-6256/140/6/1868},
	number = {6},
	urldate = {2025-07-31},
	journal = {The Astronomical Journal},
	author = {Wright, Edward L. and Eisenhardt, Peter R. M. and Mainzer, Amy K. and Ressler, Michael E. and Cutri, Roc M. and Jarrett, Thomas and Kirkpatrick, J. Davy and Padgett, Deborah and McMillan, Robert S. and Skrutskie, Michael and Stanford, S. A. and Cohen, Martin and Walker, Russell G. and Mather, John C. and Leisawitz, David and Gautier, Thomas N. and McLean, Ian and Benford, Dominic and Lonsdale, Carol J. and Blain, Andrew and Mendez, Bryan and Irace, William R. and Duval, Valerie and Liu, Fengchuan and Royer, Don and Heinrichsen, Ingolf and Howard, Joan and Shannon, Mark and Kendall, Martha and Walsh, Amy L. and Larsen, Mark and Cardon, Joel G. and Schick, Scott and Schwalm, Mark and Abid, Mohamed and Fabinsky, Beth and Naes, Larry and Tsai, Chao-Wei},
	month = dec,
	year = {2010},
	pages = {1868--1881},
}

@article{unwise,
	title = {{unWISE}: {Unblurred} {Coadds} of the {WISE} {Imaging}},
	volume = {147},
	issn = {0004-6256},
	shorttitle = {{unWISE}},
	url = {https://ui.adsabs.harvard.edu/abs/2014AJ....147..108L},
	doi = {10.1088/0004-6256/147/5/108},
	urldate = {2025-11-06},
	journal = {The Astronomical Journal},
	author = {Lang, Dustin},
	month = may,
	year = {2014},
	note = {Publisher: IOP
ADS Bibcode: 2014AJ....147..108L},
	pages = {108},
}

@article{galex,
	title = {The {Galaxy} {Evolution} {Explorer}: {A} {Space} {Ultraviolet} {Survey} {Mission}},
	volume = {619},
	issn = {0004-637X},
	shorttitle = {The {Galaxy} {Evolution} {Explorer}},
	url = {https://ui.adsabs.harvard.edu/abs/2005ApJ...619L...1M},
	doi = {10.1086/426387},
	urldate = {2025-11-05},
	journal = {The Astrophysical Journal},
	author = {Martin, D. Christopher and Fanson, James and Schiminovich, David and Morrissey, Patrick and Friedman, Peter G. and Barlow, Tom A. and Conrow, Tim and Grange, Robert and Jelinsky, Patrick N. and Milliard, Bruno and Siegmund, Oswald H. W. and Bianchi, Luciana and Byun, Yong-Ik and Donas, Jose and Forster, Karl and Heckman, Timothy M. and Lee, Young-Wook and Madore, Barry F. and Malina, Roger F. and Neff, Susan G. and Rich, R. Michael and Small, Todd and Surber, Frank and Szalay, Alex S. and Welsh, Barry and Wyder, Ted K.},
	month = jan,
	year = {2005},
	note = {Publisher: IOP
ADS Bibcode: 2005ApJ...619L...1M},
	pages = {L1--L6},
}

@misc{panstarrs,
	title = {The {Pan}-{STARRS1} {Surveys}},
	url = {https://ui.adsabs.harvard.edu/abs/2016arXiv161205560C},
	doi = {10.48550/arXiv.1612.05560},
	urldate = {2025-11-06},
	publisher = {arXiv},
	author = {Chambers, K. C. and Magnier, E. A. and Metcalfe, N. and Flewelling, H. A. and Huber, M. E. and Waters, C. Z. and Denneau, L. and Draper, P. W. and Farrow, D. and Finkbeiner, D. P. and Holmberg, C. and Koppenhoefer, J. and Price, P. A. and Rest, A. and Saglia, R. P. and Schlafly, E. F. and Smartt, S. J. and Sweeney, W. and Wainscoat, R. J. and Burgett, W. S. and Chastel, S. and Grav, T. and Heasley, J. N. and Hodapp, K. W. and Jedicke, R. and Kaiser, N. and Kudritzki, R.-P. and Luppino, G. A. and Lupton, R. H. and Monet, D. G. and Morgan, J. S. and Onaka, P. M. and Shiao, B. and Stubbs, C. W. and Tonry, J. L. and White, R. and Bañados, E. and Bell, E. F. and Bender, R. and Bernard, E. J. and Boegner, M. and Boffi, F. and Botticella, M. T. and Calamida, A. and Casertano, S. and Chen, W.-P. and Chen, X. and Cole, S. and Deacon, N. and Frenk, C. and Fitzsimmons, A. and Gezari, S. and Gibbs, V. and Goessl, C. and Goggia, T. and Gourgue, R. and Goldman, B. and Grant, P. and Grebel, E. K. and Hambly, N. C. and Hasinger, G. and Heavens, A. F. and Heckman, T. M. and Henderson, R. and Henning, T. and Holman, M. and Hopp, U. and Ip, W.-H. and Isani, S. and Jackson, M. and Keyes, C. D. and Koekemoer, A. M. and Kotak, R. and Le, D. and Liska, D. and Long, K. S. and Lucey, J. R. and Liu, M. and Martin, N. F. and Masci, G. and McLean, B. and Mindel, E. and Misra, P. and Morganson, E. and Murphy, D. N. A. and Obaika, A. and Narayan, G. and Nieto-Santisteban, M. A. and Norberg, P. and Peacock, J. A. and Pier, E. A. and Postman, M. and Primak, N. and Rae, C. and Rai, A. and Riess, A. and Riffeser, A. and Rix, H. W. and Röser, S. and Russel, R. and Rutz, L. and Schilbach, E. and Schultz, A. S. B. and Scolnic, D. and Strolger, L. and Szalay, A. and Seitz, S. and Small, E. and Smith, K. W. and Soderblom, D. R. and Taylor, P. and Thomson, R. and Taylor, A. N. and Thakar, A. R. and Thiel, J. and Thilker, D. and Unger, D. and Urata, Y. and Valenti, J. and Wagner, J. and Walder, T. and Walter, F. and Watters, S. P. and Werner, S. and Wood-Vasey, W. M. and Wyse, R.},
	month = dec,
	year = {2016},
	note = {ADS Bibcode: 2016arXiv161205560C},
}

@article{datalab,
	title = {Data {Lab}—{A} community science platform},
	volume = {33},
	issn = {22131337},
	url = {https://linkinghub.elsevier.com/retrieve/pii/S2213133720300652},
	doi = {10.1016/j.ascom.2020.100411},
	language = {en},
	urldate = {2025-01-18},
	journal = {Astronomy and Computing},
	author = {Nikutta, R. and Fitzpatrick, M. and Scott, A. and Weaver, B.A.},
	month = oct,
	year = {2020},
	pages = {100411},
}

@misc{sparcl,
	title = {{SPARCL}: {SPectra} {Analysis} and {Retrievable} {Catalog} {Lab}},
	copyright = {Creative Commons Attribution Non Commercial No Derivatives 4.0 International},
	shorttitle = {{SPARCL}},
	url = {https://arxiv.org/abs/2401.05576},
	doi = {10.48550/ARXIV.2401.05576},
	urldate = {2025-11-06},
	publisher = {arXiv},
	author = {Juneau, Stéphanie and Jacques, Alice and Pothier, Steve and Bolton, Adam S. and Weaver, Benjamin A. and Pucha, Ragadeepika and McManus, Sean and Nikutta, Robert and Olsen, Knut},
	year = {2024},
	note = {Version Number: 2},
}

@inproceedings{datalab2019,
	title = {The {NOAO} {Data} {Lab}: {Design}, {Capabilities}, and {Community} {Development}},
	volume = {523},
	shorttitle = {The {NOAO} {Data} {Lab}},
	url = {https://ui.adsabs.harvard.edu/abs/2019ASPC..523..233F},
	urldate = {2025-11-06},
	booktitle = {Astronomical {Data} {Analysis} {Software} and {Systems} {XXVII}},
	author = {Fitzpatrick, Michael and Olsen, Knut and Eychaner, Glenn and Fulmer, Leah and Huang, Lijuan and Juneau, Stephanie and Nidever, David and Nikutta, Robert and Scott, Adam},
	month = oct,
	year = {2019},
	note = {ADS Bibcode: 2019ASPC..523..233F},
	pages = {233},
}

@misc{desi,
	title = {The {DESI} {Experiment} {Part} {I}: {Science},{Targeting}, and {Survey} {Design}},
	shorttitle = {The {DESI} {Experiment} {Part} {I}},
	url = {https://ui.adsabs.harvard.edu/abs/2016arXiv161100036D},
	doi = {10.48550/arXiv.1611.00036},
	urldate = {2025-11-06},
	publisher = {arXiv},
	author = {{DESI Collaboration} and Aghamousa, Amir and Aguilar, Jessica and Ahlen, Steve and Alam, Shadab and Allen, Lori E. and Allende Prieto, Carlos and Annis, James and Bailey, Stephen and Balland, Christophe and Ballester, Otger and Baltay, Charles and Beaufore, Lucas and Bebek, Chris and Beers, Timothy C. and Bell, Eric F. and Bernal, José Luis and Besuner, Robert and Beutler, Florian and Blake, Chris and Bleuler, Hannes and Blomqvist, Michael and Blum, Robert and Bolton, Adam S. and Briceno, Cesar and Brooks, David and Brownstein, Joel R. and Buckley-Geer, Elizabeth and Burden, Angela and Burtin, Etienne and Busca, Nicolas G. and Cahn, Robert N. and Cai, Yan-Chuan and Cardiel-Sas, Laia and Carlberg, Raymond G. and Carton, Pierre-Henri and Casas, Ricard and Castander, Francisco J. and Cervantes-Cota, Jorge L. and Claybaugh, Todd M. and Close, Madeline and Coker, Carl T. and Cole, Shaun and Comparat, Johan and Cooper, Andrew P. and Cousinou, M. -C. and Crocce, Martin and Cuby, Jean-Gabriel and Cunningham, Daniel P. and Davis, Tamara M. and Dawson, Kyle S. and de la Macorra, Axel and De Vicente, Juan and Delubac, Timothée and Derwent, Mark and Dey, Arjun and Dhungana, Govinda and Ding, Zhejie and Doel, Peter and Duan, Yutong T. and Ealet, Anne and Edelstein, Jerry and Eftekharzadeh, Sarah and Eisenstein, Daniel J. and Elliott, Ann and Escoffier, Stéphanie and Evatt, Matthew and Fagrelius, Parker and Fan, Xiaohui and Fanning, Kevin and Farahi, Arya and Farihi, Jay and Favole, Ginevra and Feng, Yu and Fernandez, Enrique and Findlay, Joseph R. and Finkbeiner, Douglas P. and Fitzpatrick, Michael J. and Flaugher, Brenna and Flender, Samuel and Font-Ribera, Andreu and Forero-Romero, Jaime E. and Fosalba, Pablo and Frenk, Carlos S. and Fumagalli, Michele and Gaensicke, Boris T. and Gallo, Giuseppe and Garcia-Bellido, Juan and Gaztanaga, Enrique and Pietro Gentile Fusillo, Nicola and Gerard, Terry and Gershkovich, Irena and Giannantonio, Tommaso and Gillet, Denis and Gonzalez-de-Rivera, Guillermo and Gonzalez-Perez, Violeta and Gott, Shelby and Graur, Or and Gutierrez, Gaston and Guy, Julien and Habib, Salman and Heetderks, Henry and Heetderks, Ian and Heitmann, Katrin and Hellwing, Wojciech A. and Herrera, David A. and Ho, Shirley and Holland, Stephen and Honscheid, Klaus and Huff, Eric and Hutchinson, Timothy A. and Huterer, Dragan and Hwang, Ho Seong and Illa Laguna, Joseph Maria and Ishikawa, Yuzo and Jacobs, Dianna and Jeffrey, Niall and Jelinsky, Patrick and Jennings, Elise and Jiang, Linhua and Jimenez, Jorge and Johnson, Jennifer and Joyce, Richard and Jullo, Eric and Juneau, Stéphanie and Kama, Sami and Karcher, Armin and Karkar, Sonia and Kehoe, Robert and Kennamer, Noble and Kent, Stephen and Kilbinger, Martin and Kim, Alex G. and Kirkby, David and Kisner, Theodore and Kitanidis, Ellie and Kneib, Jean-Paul and Koposov, Sergey and Kovacs, Eve and Koyama, Kazuya and Kremin, Anthony and Kron, Richard and Kronig, Luzius and Kueter-Young, Andrea and Lacey, Cedric G. and Lafever, Robin and Lahav, Ofer and Lambert, Andrew and Lampton, Michael and Landriau, Martin and Lang, Dustin and Lauer, Tod R. and Le Goff, Jean-Marc and Le Guillou, Laurent and Le Van Suu, Auguste and Lee, Jae Hyeon and Lee, Su-Jeong and Leitner, Daniela and Lesser, Michael and Levi, Michael E. and L'Huillier, Benjamin and Li, Baojiu and Liang, Ming and Lin, Huan and Linder, Eric and Loebman, Sarah R. and Lukić, Zarija and Ma, Jun and MacCrann, Niall and Magneville, Christophe and Makarem, Laleh and Manera, Marc and Manser, Christopher J. and Marshall, Robert and Martini, Paul and Massey, Richard and Matheson, Thomas and McCauley, Jeremy and McDonald, Patrick and McGreer, Ian D. and Meisner, Aaron and Metcalfe, Nigel and Miller, Timothy N. and Miquel, Ramon and Moustakas, John and Myers, Adam and Naik, Milind and Newman, Jeffrey A. and Nichol, Robert C. and Nicola, Andrina and Nicolati da Costa, Luiz and Nie, Jundan and Niz, Gustavo and Norberg, Peder and Nord, Brian and Norman, Dara and Nugent, Peter and O'Brien, Thomas and Oh, Minji and Olsen, Knut A. G.},
	month = oct,
	year = {2016},
	note = {ADS Bibcode: 2016arXiv161100036D},
}

@article{nair2010,
	title = {A catalog of detailed visual morphological classifications for 14,034 galaxies in the sloan digital sky survey},
	volume = {186},
	issn = {00670049},
	url = {https://iopscience.iop.org/article/10.1088/0067-0049/186/2/427},
	doi = {10.1088/0067-0049/186/2/427},
	number = {2},
	urldate = {2020-12-26},
	journal = {Astrophysical Journal, Supplement Series},
	author = {Nair, Preethi B. and Abraham, Roberto G.},
	year = {2010},
	note = {arXiv: 1001.2401
Publisher: IOP Publishing},
	pages = {427--456},
}

@article{zakova2025,
	title = {Client-{Side} {Web} {Application} for {Block} {Diagram} {Processing}},
	volume = {59},
	issn = {24058963},
	url = {https://linkinghub.elsevier.com/retrieve/pii/S2405896325005968},
	doi = {10.1016/j.ifacol.2025.08.027},
	language = {en},
	number = {7},
	urldate = {2025-12-01},
	journal = {IFAC-PapersOnLine},
	author = {Žáková, Katarína and Beňačka, Cyril},
	year = {2025},
	pages = {83--86},
}
\bibliographystyle{aasjournalv7}



\end{document}